\documentclass[11pt]{article}
\usepackage[letterpaper, left=1.25in, right=1.25in, top=1.25in, bottom=1.25in]{geometry}
\usepackage[rm,bf]{titlesec}
\usepackage{enumitem}

\usepackage{amsmath}
\usepackage{amsthm}
\usepackage{multirow}
\usepackage{booktabs}

\usepackage{graphicx}
\usepackage{amssymb}
\usepackage{amsfonts}
\usepackage{amsmath,amsthm}
\usepackage[footnotesize,bf]{caption}
\usepackage{natbib}
\usepackage{setspace}
\usepackage[dvipsnames,usenames]{color}
\usepackage{pstricks,pstricks-add}
\usepackage{url}
\usepackage[usenames]{color}
\usepackage[breaklinks, colorlinks, citecolor=blue, linkcolor=blue, urlcolor=blue]{hyperref}
\usepackage{lscape}

\newtheorem{theorem}{Theorem}

\newtheorem{theorem*}[theorem]{Theorem*}

\newtheorem{definition}{Definition}
\newtheorem{definition*}[definition]{Definition*}

\newtheorem{example*}[example]{Example*}

\newtheorem{lemma}{Lemma}
\newtheorem{lemma*}[lemma]{Lemma*}

\newtheorem{proposition}{Proposition}
\newtheorem{proposition*}[proposition]{Proposition*}
\newtheorem{remark}{Remark}

\newcommand{\N}{\mathbb{N}}
\newcommand{\R}{\mathbb{R}}

\newpsobject{grilla}{psgrid}{subgriddiv=1,griddots=10,gridlabels=6pt}

\begin{document}
\bibliographystyle{elsart-harv}
\title{Empirical bias of extreme-price auctions: analysis\footnote{This paper benefited from comments of Dustin Beckett, Ernan Haruvy, Thomas Palfrey, and Anastasia Zervou as well as audiences at TETC17, U. Maryland, U. Virginia, Caltech, UC Santa Barbara, U. Rochester, Kellogg (MEDS), 2018 Naples Workshop on Equilibrium Analysis, SAET18, SCW18.  The results in this paper circulated in a paper entitled Empirical Equilibrium, which is now dedicated to study the foundations of this refinement \citep{Velez-Brown-2018-EE}. All errors are our own.}}
\date{\today}

\author{Rodrigo A. Velez\thanks{
\href{mailto:rvelezca@tamu.edu}{rvelezca@tamu.edu}; \href{https://sites.google.com/site/rodrigoavelezswebpage/home}{https://sites.google.com/site/rodrigoavelezswebpage/home}}\ \ and Alexander L. Brown\thanks{
 \href{mailto:alexbrown@tamu.edu}{alexbrown@tamu.edu}; \href{http://people.tamu.edu/\%7Ealexbrown}{http://people.tamu.edu/$\sim$alexbrown}} \\\small{\textit{Department of
Economics, Texas A\&M University, College Station, TX 77843}}}
\maketitle

\begin{abstract}
We advance empirical equilibrium analysis \citep{Velez-Brown-2019-SP} of the 
winner-bid and loser-bid auctions for the dissolution of a partnership. We show, in a complete information environment, that even though these auctions are essentially equivalent for the Nash equilibrium prediction, they can be expected to differ in fundamental ways when they are operated. Besides the direct policy implications, two general consequences follow. First, a mechanism designer who accounts for the empirical plausibility of equilibria may not be constrained by 
Maskin invariance. Second, a mechanism designer who does not account for the empirical plausibility of equilibria may inadvertently design biased mechanisms.
\begin{singlespace}

\medskip

\medskip

\end{singlespace}
\end{abstract}

\section{Introduction}\label{Sec:intro}

This paper studies the plausibility of Nash equilibria of two popular auctions for the dissolution of a partnership. By using the recently introduced empirical equilibrium analysis \citep{Velez-Brown-2019-SP}, we show that even though the Nash equilibria of the complete information games that ensue when these mechanisms are operated are essentially identical, these mechanisms can be expected to differ in fundamental ways. Two general conclusions for the theory of full implementation follow. First, a mechanism designer who accounts for empirical plausibility of equilibria may not be constrained by usual invariance properties. Second, a mechanism designer who does not account for the empirical plausibility of equilibria may design biased mechanisms.

We consider a symmetric partnership dissolution problem in which two agents who collectively own an object need to decide who receives the object when monetary compensation, chosen out of a finite but fine greed, is possible. In the spirit of the implementation literature with complete information \citep[see][for a survey]{Jackson-2001} we assume that an arbitrator who makes a recommendation for this division knows that the agents know each other well but does not know the agents' preferences on the possible divisions.\footnote{The assumption of complete information in our partnership dissolution problem allows us to contrast empirical equilibrium analysis with the well-understood restrictions of Nash implementation in this environment. Our approach to implementation theory, and in general mechanism design, does not impose restrictions on the information structure that the modeler believes is a reasonable description of reality. For instance, one can define and study empirical equilibria in general incomplete information environments, under a Bayesian information structure \citep{Velez-Brown-2019-SP}.}  This is a relevant benchmark for the dissolution of a marriage or a long standing partnership.  We assume that agents are expected utility maximizers with quasi-linear utility indices. For concreteness, suppose that agents' values for the object are $v_l\leq v_h$, which for simplicity we assume to be even numbers.   We study two prominent mechanisms that we refer to as the \textit{extreme-price auctions} (EPs). These mechanisms operate as follows. First, the arbitrator asks the agents to bid for the object. Then assigns the object to a higher bidder, breaking ties uniformly at random. The transfer from the agent who receives the object to the other agent is determined as follows: the transfer is the winner bid, the \textit{winner-bid auction} (WB); the transfer is the loser bid, the \textit{loser-bid auction} (LB).\footnote{EPs belong to the family of $\alpha$-auctions studied by \citet{Cramton-et-al-1987-Eca} and \cite{preston-1992}. Note that the partnership dissolution environment differs substantially from a buyer-seller environment in which the payoff of the loser of a first-price auction or a second-price auction does not depend on the price paid by the winner.}

Our first step to evaluate the performance of EPs is to characterize the Nash equilibrium outcomes of the games that ensue when these mechanisms are operated.  We consider both pure and mixed strategy equilibria. When agents have equal types, both auctions give, in each Nash equilibrium, equal expected payoff to each agent (Lemma~\ref{lem:charc-same-values-eq}). When valuations are different, the set of efficient Nash equilibrium payoffs of both auctions coincide (Proposition~\ref{Prop:charc-equili-auctions}).\footnote{The maximal aggregate loss in an inefficient equilibrium is one unit. The analysis of inefficient equilibria leads to the same conclusions we state here in the introduction for efficient equilibria (Sec.~\ref{Sec:results-EE-charact}).} This common set can be placed in a one to one correspondence with the integers $\{v_l/2,v_l/2+1,....,v_h/2\}$. Each equilibrium has a unique payoff-determinant bid in this set and for each such integer there is an equilibrium with this payoff-determinant bid (Proposition~\ref{Prop:Nash-payoffs}). We refer to this set as the Nash range. Between two elements of the Nash range, the higher valuation agent prefers the left one (paying less), and the lower valuation agent prefers the right one (being paid more).

Not all Nash equilibria are plausible to be observed, even approximately, when a game takes place \citep[c.f.][]{VanDamme-1991-Springer}. Thus, the equivalence of Nash equilibrium outcomes for EPs may not be realistic. If one refines the Nash equilibria of EPs by ruling out weakly dominated behavior, the associated equilibrium payoff-determinant bids collapse to extreme singleton sets: $\{v_l/2\}$ for WB and $\{v_h/2\}$ for LB. Experiments show that WB exhibits a bias in favor of the higher valuation agent and  LB exhibits a bias in favor of the lower valuation agent \citep{Brown-Velez-2017-OAU}. However, observed biases are not as extreme as those predicted by undominated equilibrium. Indeed, the best performance of these auctions, when behavior of both types clearly separates, involves persistent weakly dominated behavior \citep[Sec.~6,][] {Brown-Velez-2017-OAU}. This is consistent with the extensive literature that shows weakly dominated behavior in dominant strategy games is persistently observed \citep[see][]{Velez-Brown-2019-SP}.

Thus, we are left with the question whether the bias observed in laboratory experiments with these mechanisms should be expected in general.  That is, is there any reason why the weakly dominated behavior that sustains payoff-determinant bids close to $v_l/2$ for WB is more plausible than the weakly dominated behavior that sustains payoff-determinant bids on the other extreme of the Nash range? Symmetrically, is there any reason why the weakly dominated behavior that sustains payoff-determinant bids close to $v_h/2$ for LB is more plausible than the weakly dominated behavior that sustains payoff-determinant bids on the other extreme of the Nash range?

To answer this question we advance empirical equilibrium analysis of these auctions. This theory was introduced by \citet{Velez-Brown-2019-SP} to discriminate among weakly dominated Nash equilibria of a normal form game based on their empirical plausibility.\footnote{Traditional equilibrium refinements, implicitly of explicitly rule out all weakly dominated behavior \citep{VanDamme-1991-Springer}.  Thus, these theories are silent about the nature of the phenomenon of our interest.}

Empirical equilibrium is defined by means of the following thought experiment. Consider a researcher who samples behavior in normal-form games and constructs a theory that explains this behavior. The researcher determines the plausibility of Nash equilibria based on the empirical content of the theory by requiring that Nash equilibria be in its closure. That is, if a Nash equilibrium cannot be approximated to an arbitrary degree by the empirical content of the researcher's theory, it is identified as implausible or unlikely to be observed. Empirical equilibrium is the refinement so defined based on the non-parametric theory that each agent chooses actions with higher probability only when they are better for her given what the other agents are doing. More precisely, an empirical equilibrium is a Nash equilibrium that is the limit of a sequence of behavior satisfying \emph{weak payoff monotonicity}, i.e., between two alternative actions for an agent, say $a$ and $b$, if the agent plays $a$ with higher frequency than $b$, it is because given what the other agents are doing, $a$ has higher expected utility than $b$. Remarkably, this property is a common factor in some of the most popular models that have been proposed to account for observed behavior in experiments.\footnote{These models include the exchangeable randomly perturbed payoff models \citep{Harsanyi-1973-IJGT,VanDamme-1991-Springer}, the control cost model \citep{VanDamme-1991-Springer},  the structural QRE model \citep{mckelvey:95geb},  the regular QRE models \citep{Mackelvey-Palfrey-1996-JER,Goeree-Holt-Palfrey-2005-EE}. The common characteristic of these models that make them suitable the purpose of equilibrium refinement is that their parametric forms are independent of the game in which they are applied, and have been successful in replicating comparative statics in a diversity of games \citep{Goeree-Holt-Palfrey-2005-EE}.}

One can give  empirical equilibrium a static or dynamic interpretation. First, its definition simply articulates the logical implication of the hypothesis that the researcher's theory is well specified, for this hypothesis is refuted by the observation of a Nash equilibrium that is not an empirical equilibrium. Alternatively, suppose that the researcher hypothesizes that behavior will converge to a Nash equilibrium through an unmodeled evolutionary process that produces a path of behavior that is consistent with her theory. Then, the researcher can also conclude that the only Nash equilibria that will be approximated by behavior are empirical equilibria.

In our main results, we characterize the set of empirical equilibrium payoffs of EPs (Theorems~\ref{Thm:bias1} and~\ref{Thm:bias2}). The highlights of these characterizations are the following. With a single exception among all type profiles, these sets are disjoint. The empirical equilibrium payoffs of WB belong to the left half of the Nash range. When $v_l$ is not too close to the minimal bid, i.e., at least $3v_h/8$, the empirical equilibrium payoffs of WB essentially (up to rounding) are the left fifth of the Nash range. Symmetrically, the empirical equilibrium payoffs of LB belong to the right half of the Nash range. When $v_h$ is not too close to the maximal bid, the empirical equilibrium payoffs of LB essentially are the right fifth of the Nash range.

Thus, empirical equilibrium analysis reveals that the bias observed in experiments is indeed a characteristic of EPs. Under the testable hypotheses that observable behavior is weakly payoff monotone and that frequencies of play approach mutual best responses, WB necessarily benefits the higher valuation agent and LB necessarily benefits the lower valuation agent. This result has two policy relevant implications. First, a mechanism designer who accounts for the empirical plausibility of equilibria may not be constrained by 
Maskin invariance, a well-known necessary condition for implementation in pure strategy Nash equilibrium. Second, a mechanism designer who does not account for the empirical plausibility of equilibria may inadvertently design biased mechanisms.

The remainder of the paper is organized as follows. Sec.~\ref{Sec-Literature} places our contribution in the context of the literature. Sec.~\ref{Sec:model} introduces the model. Section~\ref{Sec:Results} presents our results, the intuition behind them, and their general implications for the theory of full implementation. Sec.~\ref{Sec:discussion} discusses the robustness of our results to violations of weak payoff monotonicity. Sec.~\ref{Sec:summary} concludes. We present all proofs in the Appendix.

\section{Related literature}\label{Sec-Literature}

EPs are part of a family of mechanisms proposed for \citet{Cramton-et-al-1987-Eca} for the dissolution of a partnership. If ownership is symmetric, in an independent private values environment, these mechanisms are efficient, individually rational, and incentive compatible. In a complete information environment, EPs are strategically equivalent to a mechanism that elicits agents valuations and then assigns an allocation in which no agent prefers the allotment of any other  agent to her own \citep[see][for a survey]{Velez-2017-Survey}.

Our study shares some features with the literature on Nash implementation \citep{Jackson-2001}, Bayesian implementation \citep{Jackson-1991-Eca}, and undominated implementation \citep{Palfrey-Srivastava-1991-Eca,Jackson-REStud-1992}. Similar  to these studies, we evaluate the performance of EPs based on the set of \textit{all} equilibrium outcomes that cannot be ruled out when the mechanisms are operated. An obvious difference is that our equilibrium refinement is a compromise between these approaches. We acknowledge that not all Nash equilibria are plausible. We do not rule out all weakly dominated equilibria, which is not supported by data.  At a technical level, our approach differs substantially with most of previous implementation literature because we consider all Nash equilibria including equilibria in mixed strategies. To our knowledge the only study that, like ours, also treats pure and mixed strategy equilibria symmetrically in the context of implementation is \citet{MEZZETTI-RENOU-2012}. These authors show that complete information implementation in Nash equilibrium (in both pure and mixed strategies) is more permissive than implementation in pure strategy Nash equilibria. Interestingly, while a continuum of mixed strategy equilibria of EPs pass our equilibrium refinement, most of its pure strategy equilibria are wiped out by it (see Sec.~\ref{Sec-implementation}). Thus, our work reinforces the view of these authors that mixed strategy equilibria and pure strategy equilibria should be treated symmetrically by the mechanism designer.  At a technical level, even though we arrive to a similar conclusion by considering plausibility of equilibria, our results are independent. The reason \citet{MEZZETTI-RENOU-2012}'s implementation in pure and mixed Nash equilibria is not bound by Maskin invariance, is that a mixed strategy equilibrium may have multiple outcomes and Maskin invariance is a requirement centered at a particular outcome. More precisely, starting from a profile of types in which a certain outcome results in a mixed strategy equilibrium, when this outcome does not drop in the ranking of any agent, the profile of strategies may not be a Nash equilibrium anymore depending on what happened to the ranking of the other outcomes in the equilibrium. Thus, the essential insight of \citet{MEZZETTI-RENOU-2012} is that one can avoid the restrictions of this property by designing mechanisms that obtain some desirable outcomes for some realizations of types only in mixed strategy equilibria with multiple outcomes. Our main insight is that a mechanism designer that accounts for plausibility of equilibria is not bound by Maskin invariance because starting from a profile of types in which a certain outcome results in an empirical equilibrium, when this outcome does not drop in the ranking of any agent, the alleged equilibrium may still be a Nash equilibrium, but may not pass our refinement anymore (see Sec.~\ref{Sec-implementation}). To provide clarity about the significance of our results, we show that the normative objectives achieved by the empirical equilibrium correspondence of EPs are not achieved by the Nash equilibrium correspondence (including equilibria in mixed strategies) of any mechanism  (Lemma~\ref{Lem:limits-marriage}).

Empirical equilibrium refines Nash equilibrium by means of approximation by behavior that is observationally sub-optimal. In this sense our work can be related with the growing literature on mechanism design with (as if) boundedly rational agents. These studies generally endorse a form of bounded rationality and aim at finding institutions that perform well when operated on agents who exhibit these particular patterns of behavior \citep[c.f.,][]{Anderson-et-al-1998-JPE,ANDERSON-et-al-2001-GEB,Eliaz-2002-RStud,HEALY-2006-JET,Cabrales2012,TUMENNASAN-2013-GEB,de-Clippel-2014-AER,de-Clippel-et-al-2017-levelk,Kneeland-2017-levelk}. Empirical equilibrium analysis does not give up the regularity induced by expected utility maximization. It puts a reality check on it by requiring proximity by plausible boundedly rational behavior. Alternatively, one can think of our analysis as the study of boundedly rational behavior when it is disciplined by proximity to a Nash equilibrium. The studies that are closest to ours consider forms of implementation that like empirical equilibrium are determined by convergence processes: \citet{Anderson-et-al-1998-JPE,ANDERSON-et-al-2001-GEB} study limit behavior of logistic QRE models in all-pay auctions and minimum contribution games; \citet{HEALY-2006-JET,Cabrales2012} study implementation in the limits of some evolutionary processes; and \citet{TUMENNASAN-2013-GEB} studies implementation in the limits of logistic equilibria for increasingly sophisticated logistic quantal response functions. \citet{Anderson-et-al-1998-JPE,ANDERSON-et-al-2001-GEB}'s approach is limited to the parametric logistic form and intensively uses the structure of the games it analyses in order to characterize logistic equilibria by means of tractable differential equations.\footnote{The set of empirical equilibria may differ from the set of Nash equilibria that are limits of logistic equilibria \citep{Velez-Brown-2018-EE}.}  \citet{HEALY-2006-JET}, \citet{Cabrales2012}, and \citet{TUMENNASAN-2013-GEB} concentrate only on pure strategy equilibria and require strong conditions be satisfied for convergence, which limit the applicability of their results to general games. 

Finally, this paper makes part of a research agenda that reevaluates the applications of game theory based on the empirical equilibrium refinement. In \citet{Velez-Brown-2018-EE} we study the foundations of empirical equilibrium. In \citet{Velez-Brown-2019-SP} we advance empirical equilibrium analysis of strategy-proof mechanisms. In a companion paper, \citet{Brown-Velez-2017-OAU}, we experimentally evaluate the predictions of empirical equilibrium for EPs. In experiments with different valuation structures these auctions exhibit the bias predicted by empirical equilibrium analysis. The data also supports the positive association of bids and their expected utility.

\section{Model}\label{Sec:model}

There are two agents $N:=\{1,2\}$ who collectively own an object (indivisible good) and need to allocate it to one of them. Monetary compensation is possible.  Each agent's payoff type is characterized by the value that he or she assigns to the object. We assume these type spaces are $\Theta_1=\Theta_2:=\{\underline{v},\underline{v}+2,...,\overline{v}\}$, where $2<\underline{v}\leq \overline{v}$ are even positive integers.\footnote{Our restrictions on valuations are only for simplicity in the presentation. The conclusions we draw from the analysis of this restricted model are general.} The generic type of agent $i$ is $v_i\in \Theta_i$. Let $\Theta:=\Theta_1\times\Theta_2$ with generic element $v:=(v_1,v_2)$. The lower and higher values at $v$ are $v_l$ and $v_h$ respectively. We also assume that agents are expected utility maximizers. The expected utility index of agent~$i$ with type $v_i$ is $v_i-p$ if receiving the object and paying $p$ to the other agent; $p$ if being paid this amount by the other agent and receiving no object. Whenever we make statements in which the identity of the agents is not relevant, we conveniently use neutral notation $i$ and $-i$. The set of possible allocations is that in which an agent receives the object and transfers an amount $p\in\{0,1,...,\overline{p}\}$ with $\overline{p}\geq\overline{v}/2+2$, to the other agent. Let $A$ be the space of these allocations. For an allocation $a\in A$ the value of agent $i$'s utility index at the allotment assigned by $a$ to this agent is $u_i(a|v_i)$.

A social choice correspondence (scc) selects a set of allocations for each possible profile of types. The generic scc is $G:\Theta\rightrightarrows A$.

A mechanism is a pair $(M,\varphi)$ where $M:=(M_i)_{i\in N}$ is an unrestricted message space and $\varphi:M\rightarrow \Delta(A)$ is an outcome function. Our main results concern the finite mechanisms we introduce next. For simplicity we complete the set up of the model assuming that mechanisms are finite. Lemma~\ref{Lem:limits-marriage}, our only result that refers to general mechanisms, can be generalized for appropriate extensions of strategies and Nash equilibrium.

\begin{definition}\rm The \textit{winner-bid auction} is the mechanism in which each agent selects a bid in the set $\{0,1,2,...,\overline{p}\}$. An agent with the highest bid receives the object. Ties are resolved uniformly at random. The agent who receives the object pays the winner bid to the other agent. The \textit{loser-bid auction} is the mechanism defined similarly where the payment is the loser bid. We refer to these two auctions as the \textit{extreme-price auctions}.
\end{definition}

Given a profile of types $v\in\Theta$, mechanism $(M,\varphi)$ determines a complete information game $\Gamma:=(M,\varphi,v)$. A (mixed) strategy for agent~$i$ is a probability measure on $M_i$. Agent~$i$'s generic strategy is $\sigma_i\in\Delta(M_i)$.  The profile of strategies is $\sigma:=(\sigma_i)_{i\in N}$. A pure strategy places probability one in a given action. We associate pure strategies with actions themselves. The expected utility of agent~$i$ with type $v_i$ in $\Gamma$ from selecting action $m_i$  when the other agent selects an action as prescribed by $\sigma_{-i}$ is
\[U_\varphi(m_i|\sigma_{-i};v_i):= \sum_{m_{-i}\in M_{-i}} u_i(\varphi(m_i,m_{-i})|v_i)\sigma_{-i}(m_{-i}).\]

A profile of strategies $\sigma$  is a \textit{Nash equilibrium} of $\Gamma$ if for each $i\in N$, each $m_i$ in the support of $\sigma_i$, and each $m_i'\in M_i$, $U_\varphi(m_i'|\sigma_{-i};v_i)\leq U_\varphi(m_i|\sigma_{-i};v_i)$. The set of Nash equilibria of $\Gamma$ is $N(\Gamma)$. The set of Nash equilibrium outcomes of $\Gamma$ are those obtained with positive probability for some Nash equilibrium of $\Gamma$. Agent~$i$'s expected payoff in equilibrium $\sigma$ is $\pi_i(\sigma)$, where for simplicity in the notation we avoid conditioning these payoffs on the agent's types, which can be always inferred from context.

\begin{definition}[\citealp{Velez-Brown-2018-EE}]\rm $\sigma:=(\sigma_i)_{i\in N}$ is \textit{weakly payoff monotone for $(M,\varphi,v)$} if for each $i\in N$ and each pair $\{m_i,n_i\}\subseteq M_i$, $\sigma_i(m_i)>\sigma_i(n_i)$ implies that $U_\varphi(m_i|\sigma_{-i};v_i)> U_\varphi(n_i|\sigma_{-i};v_i)$.\end{definition}

\begin{definition}[\citealp{Velez-Brown-2018-EE}]\rm $\sigma\in N(\Gamma)$ is an \textit{empirical equilibrium of $\Gamma$} if there is a sequence of weakly payoff monotone distributions of $\Gamma$, $\{\sigma^\lambda\}_{\lambda\in\N}$, such that as $\lambda\rightarrow\infty$, $\sigma^\lambda\rightarrow\sigma$.\end{definition}

\section{Results}\label{Sec:Results}

\subsection{Extreme-price auctions and Nash equilibrium}\label{Sec:results-NE}

We start by characterizing the Nash equilibrium outcomes of the games induced by the EPs.

The theory of fair allocation has produced a series of principles that an arbitrator may want to adhere to when resolving a partnership dissolution dispute \citep[see][for a survey]{Thomson-2006}. Two of the most popular are the following: \textit{efficiency}, i.e., a party who values the object the most should receive it, or equivalently, that expected payoffs add up to $v_h$; and \textit{equity}, i.e., no party should prefer the allotment of the other \citep{Foley-1967-YEE}. Abstracting from incentives issues imagine that the arbitrator knows the agents' valuations, $v\in\Theta$. It is easy to see that if the arbitrator abides by the principles of efficiency and equity, agents' payoffs should have the form:\footnote{In our environment no-envy implies efficiency for deterministic allocations \citep{Svensson-1983-Eca}. Our statement here refers also to random assignments.} $v_l/2+t$ for $l$ and $v_h/2+(|v_h/2-v_l/2|-t)$ for $h$, where $0\leq t\leq |v_h/2-v_l/2|$. In other words, if an arbitrator endorses these two principles, the only that is left for him or her is to determine a division between the agents of the 
``equity surplus,'' i.e., $ES(v):=|v_h/2-v_l/2|$ \citep{Tadenuma-and-Thomson-1995-TD}.

The following lemma states that each extreme-price auction achieves, in each Nash equilibrium (including those in mixed strategies), the objectives of efficiency and equity when agents have equal valuations. We omit the straightforward proof.
\begin{lemma}\rm\label{lem:charc-same-values-eq}Let $(M,\varphi)$ be an extreme-price auction and $v\in\Theta$ such that $v_1=v_2$. Then for each $\sigma\in N(M,\varphi,v)$, $\pi_1(\sigma)=\pi_2(\sigma)=v_1/2=v_2/2$. Moreover, for each deterministic allocation $a\in A$ in which  each agent receives this common payoff, there is a pure-strategy Nash equilibrium of $(M,\varphi,v)$ whose outcome is this allocation.
\end{lemma}
The following proposition states that each extreme-price auction essentially achieves, in each Nash equilibrium (including those in mixed strategies), the objectives of efficiency and equity for arbitrary valuation profiles.

\begin{proposition}\rm\label{Prop:charc-equili-auctions}Let $(M,\varphi)$ be an extreme-price auction and $v\in\Theta$ such that $v_l<v_h$. Let $\sigma$ be a Nash equilibrium of $(M,\varphi,v)$. Then, there is $p\in\{v_l/2,...,v_h/2\}$ such that the support of $\sigma_{l}$ belongs to $\{0,...,p\}$ and the support of $\sigma_{h}$ belongs to $\{p,...,\overline{p}\}$. If $\sigma$ is efficient, the higher value agent receives the object and pays $p$ to the other agent. Moreover,
    \begin{enumerate}
    \item If $(M,\varphi)$ is the winner-bid auction, then $p$ is in the support of $\sigma_h$. If $\sigma$ is inefficient, i.e., $\sigma_l(p)>0$, then $p=v_l/2$; $\pi_l(\sigma)+\pi_h(\sigma)\geq v_h-1$;  and $\pi_h(\sigma)\geq v_h/2+ES(v)-1$.

    \item If $(M,\varphi)$ is the loser-bid auction, then $p$ is in the support of $\sigma_l$. If $\sigma$ is inefficient, i.e., $\sigma_h(p)>0$, then $p=v_m/2$;  $\pi_l(\sigma)+\pi_h(\sigma)\geq v_h-1$; and $\pi_l(\sigma)\geq v_l/2+ES(v)-1$.
        \end{enumerate}
\end{proposition}

Proposition~\ref{Prop:charc-equili-auctions} reveals that the Nash equilibria of the extreme-price auctions have a simple structure. In all equilibria, the payoff-determinant bid is in the set $\{v_l/2,...,v_h/2\}$.  Let us refer to this set of bids as the Nash range. In most of these equilibria agents' bids are strictly separated. That is, there is a bid $p$ in the Nash range such that one agent bids weakly on one side of $p$ and the other agent bids strictly on the other side of $p$. In these equilibria, which are strictly separated, outcomes are efficient and equitable. There are inefficient equilibria. For the winner-bid auction, it is possible that both agents bid $v_l/2$. For the loser-bid auction it is possible that both agents bid $v_h/2$. In both cases the aggregate welfare loss is at most one unit, i.e., the size of the minimal difference between bids.  This means that if the minimal bid increment is one cent, the maximum that these auctions can lose in aggregate expected utility for any Nash equilibrium is one cent. Thus, one can say that these auctions essentially implement the principles of efficiency and equity in Nash equilibria.\footnote{Our proof of Proposition~\ref{Prop:charc-equili-auctions} also reveals that the probability of an inefficient outcome is bounded above by the inverse of the equity surplus, measured in the minimal bid increment.}

Proposition~\ref{Prop:charc-equili-auctions} allows us to characterize Nash equilibrium payoffs for extreme-price auctions. For efficient equilibria, these payoffs are exactly the integer divisions of the equity surplus. For simplicity in the presentation, in the remainder of the paper we omit the analysis of inefficient equilibria. We know from Proposition~\ref{Prop:charc-equili-auctions}  that the corresponding payoffs are extreme in the equilibrium set. Thus, all conclusions we obtain for efficient empirical equilibria extend to inefficient empirical equilibria.

\begin{proposition}\rm\label{Prop:Nash-payoffs}Let $(M,\varphi)$ be an extreme-price auction and $v\in\Theta$ such that $v_l<v_h$. The set of efficient Nash equilibrium payoffs of $(M,\varphi,v)$, i.e.,\linebreak $\{(\pi_l(\sigma),\pi_{h}(\sigma)):\sigma\in N(M,\varphi,v),\pi_l(\sigma)+\pi_h(\sigma)=v_h\}$, is the set of integer divisions of the equity-surplus, i.e., $\{(v_l/2+ES(v)-t,v_h/2+t):t\in\{0,1,...,ES(v)\}\}$.
\end{proposition}

\subsection{Empirical equilibrium}\label{Sec:results-EE}

\subsubsection{Empirical equilibrium payoffs}\label{Sec:results-EE-charact} We characterize the set of empirical equilibrium payoffs of the extreme-price auctions. Recall that by Lemma~\ref{lem:charc-same-values-eq}, when valuations are equal, the payoff of each agent is the same in each Nash equilibrium of each extreme-price auction. Thus, we only need to characterize the payoffs of empirical equilibria when agents' valuations differ. Since Proposition~\ref{Prop:Nash-payoffs} characterizes Nash equilibrium payoffs for these auctions, it is convenient to describe empirical equilibrium payoffs by the set of conditions for which a Nash equilibrium payoff is an empirical equilibrium payoff.

\begin{theorem}\rm\label{Thm:bias1}Let $(M,\varphi)$ be the winner-bid auction, $v\in \Theta$ such that $v_l<v_h$, and $\sigma\in N(M,\varphi,v)$. If $\sigma$ is efficient, $\pi(\sigma)$ is the payoff of an empirical equilibrium of $(M,\varphi,v)$ if and only if
      \begin{enumerate}
        \item $\pi_h(\sigma)=v_h/2+ES(v)$ when $ES(v)\leq 2$;
        \item $\pi_h(\sigma)\geq v_h/2+ES(v)/2+v_l/4-1/2$ when $ES(v)>2$ and $v_l\leq 3v_h/8$;
        \item $\pi_h(\sigma)\geq v_h/2+4ES(v)/5-4/5$ when $ES(v)>2$, $v_l>3v_h/8$, and $v_l<7v_h/12-7/6$;
        \item $\pi_h(\sigma)>v_h/2+4ES(v)/5-4/5$ when $ES(v)>2$, $v_l>3v_h/8$, and $v_l\geq 7v_h/12-7/6$.
      \end{enumerate}
\end{theorem}

\begin{figure}[h]
\centering
\begin{pspicture}(-.2,-1)(11.5,7)
\psline{->}(-.2,0)(11.5,0)
\psline(-.15,.5)(.15,.5)
\psline{->}(0,-.2)(0,6.5)
\rput[c](6,-.5){$\mbox{\footnotesize{$v_h/2$}}$}
\rput[c](0,6.8){$\mbox{\footnotesize{$v_l/2$}}$}
\rput[c](3,-.5){$\mbox{\footnotesize{$v_h/4$}}$}
\rput[r](-.2,.5){$\mbox{\footnotesize{$1$}}$}
\psdots[dotstyle=|](2.25,0)(3,0)(6,0)
\psline{->}(1.25,-.6)(2.25,-.6)(2.25,-.1)
\psline[linestyle=dotted](2.25,0)(2.25,2.25)
\rput[r](1.2,-.6){$\mbox{\footnotesize{$(3/8)v_h/2$}}$}
\rput[c](11.5,-.5){$\mbox{\footnotesize{$p$}}$}
\psdots[dotstyle=triangle,dotsize=5pt](0.5,0.5)(1,.5)(1.5,.5)(2,.5)(2.5,.5)(3,.5)
\psdots[dotstyle=square,dotsize=5pt](3.5,.5)(4,.5)(4.5,.5)(5,.5)(5.5,.5)(6,.5)
\psdots[dotsize=2.5pt](0.5,0.5)(1,.5)(1.5,.5)(2,.5)(2.5,.5)(3,.5)(3.5,.5)(4,.5)(4.5,.5)(5,.5)(5.5,.5)
\psdot[dotsize=2.5pt](6,.5)

\psdots[dotsize=5pt,dotstyle=triangle](1,1)(1.5,1)(2,1)(2.5,1)(3,1)
\psdots[dotsize=5pt,dotstyle=square](4,1)(4.5,1)(5,1)(5.5,1)(6,1)
\psdots[dotsize=2.5pt](1,1)(1.5,1)(2,1)(2.5,1)(3,1)(3.5,1)(4,1)(4.5,1)(5,1)(5.5,1)
\psdot[dotsize=2.5pt](6,1)
\psdots[dotsize=5pt,dotstyle=triangle](1.5,1.5)(2,1.5)(2.5,1.5)(3,1.5)
\psdots[dotsize=5pt,dotstyle=square](4.5,1.5)(5,1.5)(5.5,1.5)(6,1.5)
\psdots[dotsize=2.5pt](1.5,1.5)(2,1.5)(2.5,1.5)(3,1.5)(3.5,1.5)(4,1.5)(4.5,1.5)(5,1.5)(5.5,1.5)
\psdot[dotsize=2.5pt](6,1.5)
\psdots[dotsize=5pt,dotstyle=triangle](2,2)(2.5,2)(3,2)
\psdots[dotsize=5pt,dotstyle=square](5,2)(5.5,2)(6,2)
\psdots[dotsize=2.5pt](2,2)(2.5,2)(3,2)(3.5,2)(4,2)(4.5,2)(5,2)(5.5,2)
\psdot[dotsize=2.5pt](6,2)
\psdots[dotsize=5pt,dotstyle=triangle](2.5,2.5)(3,2.5)(3.5,2.5)
\psdots[dotsize=5pt,dotstyle=square](5,2.5)(5.5,2.5)(6,2.5)
\psdots[dotsize=2.5pt](2.5,2.5)(3,2.5)(3.5,2.5)(4,2.5)(4.5,2.5)(5,2.5)(5.5,2.5)
\psdot[dotsize=2.5pt](6,2.5)
\psline[linewidth=.3pt]{|<->|}(2.5,2.7)(3.6,2.7)
\psline[linewidth=.3pt](2,3)(2.5,3)(3.05,2.7)
\rput[r](1.9,3){$\mbox{\tiny$ES(v)/5+4/5$}$}
\psdots[dotsize=5pt,dotstyle=triangle](3,3)(3.5,3)(4,3)
\psdots[dotsize=5pt,dotstyle=square](5,3)(5.5,3)(6,3)
\psdots[dotsize=2.5pt](3,3)(3.5,3)(4,3)(4.5,3)(5,3)(5.5,3)
\psdot[dotsize=2.5pt](6,3)
\psdots[dotsize=5pt,dotstyle=triangle](3.5,3.5)(4,3.5)
\psdots[dotsize=5pt,dotstyle=square](5.5,3.5)(6,3.5)
\psdots[dotsize=2.5pt](3.5,3.5)(4,3.5)(4.5,3.5)(5,3.5)(5.5,3.5)
\psdot[dotsize=2.5pt](6,3.5)
\psdots[dotsize=5pt,dotstyle=triangle](4,4)(4.5,4)
\psdots[dotsize=5pt,dotstyle=square](5.5,4)(6,4)
\psdots[dotsize=2.5pt](4,4)(4.5,4)(5,4)(5.5,4)
\psdot[dotsize=2.5pt](6,4)
\psdots[dotsize=5pt,dotstyle=triangle](4.5,4.5)(5,4.5)
\psdots[dotsize=5pt,dotstyle=square](5.5,4.5)(6,4.5)
\psdots[dotsize=2.5pt](4.5,4.5)(5,4.5)(5.5,4.5)
\psdot[dotsize=2.5pt](6,4.5)
\psdots[dotsize=5pt,dotstyle=triangle](5,5)
\psdots[dotsize=5pt,dotstyle=square](6,5)
\psdots[dotsize=2.5pt](5,5)(5.5,5)
\psdot[dotsize=2.5pt](6,5)
\psdots[dotsize=5pt,dotstyle=triangle](5.5,5.5)
\psdots[dotsize=5pt,dotstyle=square](6,5.5)
\psdot[dotsize=2.5pt](5.5,5.5)
\psdot[dotsize=2.5pt](6,5.5)
%
%
\psframe(1,6)(3.2,4)
\psdot[dotsize=2.5pt](1.3,5.75)
\rput[l](1.5,5.75){\tiny Nash range}
\psdot[dotsize=5pt,dotstyle=triangle](1.3,5.25)
\rput[l](1.5,5.25){\tiny winner-bid}
\psdot[dotsize=5pt,dotstyle=square](1.3,4.75)
\rput[l](1.5,4.75){\tiny loser-bid}
\psdot[dotsize=5pt,dotstyle=pentagon](1.3,4.25)
\rput[l](1.5,4.25){$\mbox{\tiny $\overline{p}$}$}
\psdots[dotsize=5pt,dotstyle=pentagon](6.5,.5)(7,1)(7.5,1.5)(8,2)(8.5,2.5)(9,3)(9.5,3.5)(10,4)(10.5,4.5)(11,5)(11.5,5.5)
\end{pspicture}
\caption{\textbf{Payoff determinant bids in an empirical equilibrium of the extreme-price auctions.} For a given $v\in\Theta$ such that $v_l<v_h$, the Nash range is the set $\{v_l/2,...,v_h/2\}$. In this graph we show, for a fixed $v_h$, the Nash range for each $2<v_l< v_h$ and the bids that are winning bids for empirical equilibria of the winner-bid and loser-bid auction. In the vertical line we measure $v_l/2$ and in the horizontal line we measure bids. For $v_l=2$, the Nash range is the set $\{1,2,...,v_h/2\}$. In the graph this set is represented by the points $(1,1)$, $(2,1)$,...,$(v_h/2,1)$ (black dots). In general, for $v_l$, we represent the Nash range, i.e., $\{v_l/2,...,v_h/2\}$ by the set of points $(v_l/2,v_l/2)$,...,$(v_h/2,v_l/2)$. We represent the winner bids in an empirical equilibrium of the winner-bid auction for a given $v\in \Theta$ by triangles. We represent the winner bids in an empirical equilibrium of the loser's-bid auction for a given $v\in \Theta$ such that $v_l/2=\overline{p}-v_h/2$ by squares. Notice that this choice of variable $\overline{p}$ as $v_l$ increases guarantees symmetry, i.e., there is the same number of possible bids on each side of the Nash range, and is chosen for easier interpretation of the graph.}\label{Fig:winning-bid-ext-p-auc}
\end{figure}
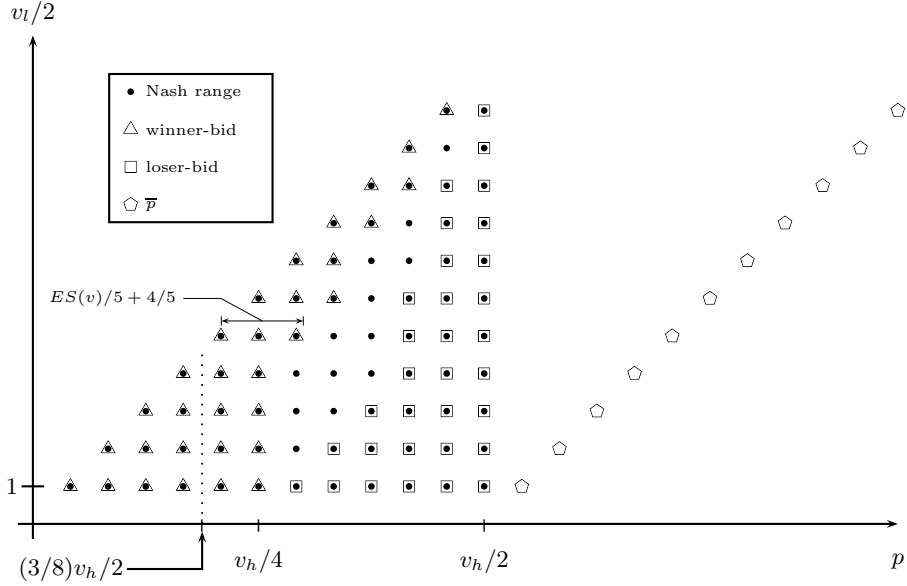

Theorem~\ref{Thm:bias1} reveals a surprising characteristic of the empircal equilibria of the winner-bid auction.  For simplicity fix $v_h$ at a certain value. Let $v_l\leq v_h$. When $v_l$ is low, i.e., at most $3v_h/8$, the minimal share of the equity surplus that the higher value agent obtains in an empirical equilibrium is, essentially, at least 50\% of the equity surplus (since we assumed $v_l$ is a positive even number, the exact share depends on rounding, but is never less than 50\%). More precisely, for this range of $v_l$, agent~$h$ receives a payoff that is at least
\[v_h/2+ES(v)/2+v_l/4-1/2=v_h/2+v_h/4-1/2.\]
This means that $p$ is the winner bid in an empirical equilibrium of the winner-bid auction for such valuations if and only if $p\leq v_h/4+1/2$ (Fig.~\ref{Fig:winning-bid-ext-p-auc}).  Thus, while the maximal bid in an empirical equilibrium is the same for all valuations when $v_l\leq 3v_h/8$, the minimal percentage of the equity surplus that is assigned to the higher value agent increases from essentially 50\% when $v_l=2$ to essentially 80\% when $v_l=3v_h/8$. For higher values of $v_l$, i.e., $3v_h/8<v_l<v_h$, the higher value agent receives, essentially, at least 80\% of the equity surplus (Fig.~\ref{Fig:winning-bid-ext-p-auc}).

In summary, Theorem~\ref{Thm:bias1} states that the minimal share of the equity surplus that the higher value agent obtains in an empirical equilibrium of the winner-bid auction depends on the number of possible bids that are to the left of the Nash range. In the extreme case in which there is only one bid to the left of the Nash range, the higher value agent essentially obtains at least 50\% of the equity surplus. As the number of bids to the left of the Nash range increases, the minimal share of the equity surplus that is obtained by the higher value agent in an empirical equilibrium increases until it reaches essentially 80\% when the number of possible bids to the left of the Nash range is 60\% of the number of bids in the Nash range (equivalently, $v_l\leq 3v_h/8$). When the number of possible bids to the left of the Nash range is higher than 60\% of the number of bids in the Nash range (equivalently, $v_l>3v_h/8$), the minimal share of the equity surplus that is obtained by the higher value agent in an empirical equilibrium remains essentially 80\%. (For low values of the equity surplus, rounding has a significant effect; see Fig.~\ref{Fig:winning-bid-ext-p-auc}).

\begin{theorem}\rm\label{Thm:bias2}Let $(M,\varphi)$ be the loser-bid auction, $v\in \Theta$ such that $v_l<v_h$, and  $\sigma\in N(M,\varphi,v)$. If $\sigma$ is efficient, $\pi(\sigma)$ is the payoff of an empirical equilibrium of $(M,\varphi,v)$ if and only if
     \begin{enumerate}
       \item $\pi_l(\sigma)\geq v_l/2+ES(v)$ if $ES(v)\leq2$;
       \item $\pi_l(\sigma)\geq v_l/2+ES(v)/2+(\overline{p}-v_h/2)/2-1/2$ if $ES(v)>2$ and $v_h/2\geq v_l/2+5(\overline{p}-v_l/2)/8$.
       \item $\pi_l(\sigma)\geq v_l/2+4ES(v)/5-4/5$ if $ES(v)>2$, $v_h/2< v_l/2+5(\overline{p}-v_l/2)/8$, and $v_h/2>\overline{p}-7(\overline{p}-v_l/2)/24-7/12$;
       \item $\pi_l(\sigma)>v_l/2+4ES(v)/5-4/5$ if $ES(v)>2$, $v_h/2< v_l/2+5(\overline{p}-v_l/2)/8$, and $v_h/2\leq \overline{p}-7(\overline{p}-v_l/2)/24-7/12$.
     \end{enumerate}
\end{theorem}

The empirical equilibrium payoffs of the loser-bid auction are symmetric to those of the winner-bid auction. The minimal share of the equity surplus that the lower value agent obtains in an empirical equilibrium of the winner-bid auction depends on the number of possible bids that are to the right of the Nash range. In the extreme case in which there is only one bid to the right of the Nash range, the lower value agent essentially obtains at least 50\% of the equity surplus. As the number of bids to the right of the Nash range increases, the minimal share of the equity surplus that is obtained by the lower value agent in an empirical equilibrium increases until it reaches essentially 80\% when the number of possible bids to the right of the Nash range is 60\% of the number of bids in the Nash range. When the number of possible bids to the right of the Nash range is higher than 60\% of the number of bids in the Nash range, the minimal share of the equity surplus that is obtained by the lower  value agent in an empirical equilibrium remains essentially 80\%. (For low values of the equity surplus, rounding has a significant effect; see Fig.~\ref{Fig:winning-bid-ext-p-auc}).

\subsubsection{The intuition}  The characterization of empirical equilibrium payoffs of the extreme-price auctions reveals that empirical equilibrium makes a delicate selection of Nash equilibria, which is sensitive to the global structure of the game. An empirical equilibrium may involve an agent playing a weakly dominated strategy. However, not all Nash equilibria, in particular not all Nash equilibria in which an agent plays a weakly dominated strategy, are empirical equilibria. Thus, empirical equilibrium somehow discriminates among actions by assessing how likely they can be played based on how they compare with the other actions. A discussion of the proof of Theorem~\ref{Thm:bias1} informs us about this feature of empirical equilibrium.

Let $v\in\Theta$ be such that $v_l<v_h$. Let $(M,\varphi)$ be the winner-bid auction and consider  $\sigma\in N(M,\varphi,v)$. By Proposition~\ref{Prop:charc-equili-auctions} this equilibrium is characterized by a winner bid $p\in\{v_l/2,...,v_h/2\}$. Suppose that $p>v_l/2$ and let $t:= p-1-v_l/2$. Suppose for simplicity that $t\geq 1$. One can easily see that agent $l$ always regrets wining the auction when bidding $p-1$. Indeed, bidding $p-1$ is weakly dominated for agent $l$ by each bid in $\{v_l/2-1,v_l/2,...,p-2\}$. Let $\sigma^\lambda$ be any monotone quantal response equilibrium of $(M,\varphi,v)$. It follows that for each $b\in \{v_l/2-1,v_l/2,...,p-2\}$, $U_\varphi(b|\sigma^\lambda_h;v_l)\geq U_\varphi({p-1}|\sigma^\lambda_h;v_l)$. This means that there are always at least $t+2$ bids that are at least as good as $p-1$ for agent $l$. This imposes a cap on the probability that this agent can place on $p-1$. More precisely,
\begin{equation}\sigma^\lambda_l(p-1)\leq 1/(t+2).\label{Eq:discussion1}\end{equation}
Thus, if $\sigma$ is an empirical equilibrium, by (\ref{Eq:discussion1}), $\sigma_l(p-1)$ needs to be the limit of a sequence of probabilities bounded above by $1/(t+2)$ and thus,
\begin{equation}\label{Eq:Discussion2}\sigma_l(p-1)\leq 1/(t+2).\end{equation}
Now, in order for $\sigma$ to be a Nash equilibrium it has to be the case that agent $h$ has no incentive to bid one unit less than $p$. This is simply,
\[v_h-p\geq \sigma_l(p-1)\left[\frac{1}{2}(v_h-(p-1))+\frac{1}{2}(p-1)\right]+(1-\sigma_l(p-1))(v_h-(p-1)).\]
Equivalently,  $\sigma_l(p-1)\geq1/(v_h/2-p+1)$. Together with (\ref{Eq:Discussion2}) this implies that $p\leq (v_l/2+v_h/2)/2$, or
equivalently
\[\pi_h(\sigma)\geq v_h/2+ES(v)/2.\]
Thus, the reason why empirical equilibrium predicts agent~$h$ gets at least half of the equity surplus for any possible valuation is that it is not plausible that agent~$l$ will consistently bid above half of the Nash range. These high bids are worse than too many bids to their left for agent~$l$. If this agent's actions are monotone with respect to utility, the maximum probability that he or she will end up placing in a bid on the right half of the Nash range will never be enough to contain the propensity of agent $h$ to lower his or her bid.

When there is more than one bid to the left of the Nash range, the analysis becomes subtler. Let $y:= v_l/2-3t$. The key to complete this analysis is to prove that in any quantal response equilibrium of $(M,\varphi,v)$ that is close to $\sigma$, all bids $b\in\{\max\{0,y\},..., p-1\}$ are weakly better than $p-1$ for agent~$l$. The subtlety here lies in that this is not implied directly by a weak domination relation as in our analysis above. In order to uncover this we need to recursively obtain estimates of agent $h$'s distribution of play, which in turn depend on agent~$l$'s distribution of play.

Remarkably, we prove that the set of restrictions that we uncover by means of our analysis are the only ones that need to be satisfied by an empirical equilibrium payoff. \citet{Goeree-Holt-Palfrey-2005-EE} characterize the set of regular QRE of an asymmetric matching pennies game. At a conceptual level, this exercise is similar to the construction of empirical equilibria. However, the techniques developed by \citet{Goeree-Holt-Palfrey-2005-EE} are useful only in two-by-two games, where an agent's distribution of play is described by a single real number. Thus, there is virtually no precedent in the construction of empirical equilibria that are not strict Nash equilibria for games with non-trivial action spaces. We do so as follows. First, we identify for each target payoff an appropriate Nash equilibrium that produces it, say~$\sigma$. Then we take a convex combination of a perturbation of~$\sigma$ and a logistic quantal response. This defines a continuous operator whose fixed points are the basis of our construction. In a two-limit process we first get close enough to~$\sigma$ by placing a weight on it that is high enough so the fixed points of the convex combination operators inherit some key properties of~$\sigma$. Then we allow the logistic response to converge to a best response and along this path of convergence we obtain interior distributions that are close to~$\sigma$ and are payoff-monotone.

\subsection{Empirical equilibrium analysis and full implementation theory}\label{Sec-implementation}

Full implementation theory is the application of game theory in which one designs mechanisms based on their worst-case scenario performance. More precisely, given an scc and a prediction for the interaction of agents in strategic situations, the designer looks for a mechanism that obtains outcomes selected by the scc for the true profile of types for \emph{all} the  predicted outcomes of the game that ensues when the mechanism is operated. When this exercise is done based on the Nash equilibrium prediction, the mechanism designer finds that many a desirable social objective is beyond her reach \citep{Maskin-1999-RES,Jackson-1991-Eca}.  By contrast, most of the constraints of the mechanism designer are lifted if the worst-case scenario discards all Nash equilibria that involve weakly dominated behavior \citep{Palfrey-Srivastava-1991-Eca,Jackson-REStud-1992}.

Since there are Nash equilibria that are intuitively implausible, designing for a worst-case scenario accounting for all possible Nash equilibria is unnecessarily pessimistic. On the other extreme, ruling out all weakly dominated behavior, is also unrealistically optimistic \citep[see][]{Velez-Brown-2019-SP}. Our partnership dissolution problem epitomizes this conflict and clearly shows how empirical equilibrium analysis resolves it.

One can envision an arbitrator having a deliberate choice on the division of the equity surplus in a partnership dissolution problem. For instance the arbitrator may want to guarantee that the lower valuation agent always receive at least half of the equity surplus. Proposition~\ref{Prop:Nash-payoffs} implies that this is not achieved by the extreme-price auctions if one evaluates them with the Nash equilibrium prediction. The following lemma states that this is a feature not only of these auctions but also of any mechanism with at least one equitable Nash equilibrium for each profile of types.

\begin{lemma}\rm\label{Lem:limits-marriage}Let $(M,\varphi)$ be a mechanism. Suppose that for each $v\in\Theta$, there is a Nash equilibrium of $(M,\varphi,v)$ (potentially in mixed strategies) that obtains with positive probability only efficient and equitable allocations for~$v$. Then, for each $v\in\Theta$ such that $v_l<v_h$, and each $\{t,t+1\}\subseteq \{0,1,...,ES(v)\}$ there is an efficient $\sigma\in N(M,\varphi,v)$ such that $v_l/2+t\leq\pi_l(\sigma)\leq v_l/2+t+1$.
\end{lemma}

Lemma~\ref{Lem:limits-marriage}  states that if an arbitrator selects a mechanism that obtains efficiency and equity in at least one Nash equilibrium (partial implementation), then the range of divisions of the equity surplus that are necessarily obtained when valuations are different is essentially the whole Nash range. More precisely, for any two consecutive integer divisions of the equity surplus, there is a Nash equilibrium that obtains a division of the equity surplus between these two divisions.

Theorems~\ref{Thm:bias1} and~\ref{Thm:bias2} and Lemma~\ref{Lem:limits-marriage} allow us to conclude that a social planner who accounts for the empirical plausibility of equilibria, realizes that some social goals, which would be ruled out impossible by the full Nash implementation analysis (independently whether implementation in mixed strategy equilibria as in \citet{MEZZETTI-RENOU-2012} is allowed or not), are within his or her reach. For instance, using LB could be sensible for a social planner who is able to exercise some level of affirmative action and chooses to benefit a segment of the population who are likely to have lower valuations for the objects to be assigned.

Finally, it is worth noting that our results suggest that it is not without loss of generality to restrict our attention to pure-strategy equilibria when a mechanism is operated.

\begin{remark}\rm \label{Rem:not-pur-strat}Let $v\in\Theta$ such that $v_l<v_h$. The only pure-strategy empirical equilibrium of the winner-bid auction for $v$ is $\sigma_l=\delta_{v_l/2-1}$ and $\sigma_h:=\delta_{v_l/2}$. The only pure-strategy empirical equilibrium of the loser-bid auction for $v$ is $\sigma_l=\delta_{v_h/2}$ and $\sigma_h:=\delta_{v_h/2+1}$.
\end{remark}

\citet{Jackson-REStud-1992} constructed examples in which including arguably plausible mixed-strategy equilibria in a worst-case scenario analysis would reverse the conclusions one obtains by only analyzing pure-strategy equilibria. Empirical equilibrium analysis goes beyond these observations and provides a clear framework in which plausibility is built into the prediction of agents' behavior. It is fair to say then that while empirical mechanism design opens new possibilities in the design of economic institutions, it also sets the standards of analysis high by forcing us to consider mixed-strategy equilibria. In this context, our complete characterization of empirical equilibria of extreme-price auctions in Theorems ~\ref{Thm:bias1} and~\ref{Thm:bias2}, and the characterization results in \citet{Velez-Brown-2019-SP}, show that the technical challenges can be resolved in policy relevant environments.

\section{Discussion}\label{Sec:discussion}

Our analysis of mechanism based on empirical equilibrium has some challenges that are shared by the analysis of mechanisms based on any solution concept that is not single-valued for each game. If a Nash equilibrium is not empirical, we learn that it is implausible (under our hypothesis of weak payoff monotonicity). By contrast, if a Nash equilibrium is empirical, we do not know whether the equilibrium will be actually relevant in a particular environment.

In some applications this may not be a problem. For instance, the main conclusion from our analysis of empirical equilibria of EPs survives even if some empirical equilibria end up not being relevant for these auctions. More precisely, the bias predicted for these mechanisms will be present even when not all empirical equilibria end up being relevant when they are operated. In general, many a social objective is guaranteed to be achieved whenever all empirical equilibrium outcomes that result when a mechanism is operated satisfy it. Think, for instance, of 
Pareto efficiency. Suppose that a mechanism guarantees this property for each empirical equilibrium.  This means that as long as behavior is payoff monotone and approximates mutual best responses, this behavior will obtain Pareto efficiency with high probability. Again, notice that in order to conclude this, it is not necessary to know which empirical equilibria will actually be relevant when the mechanism is operated.

Designing based on empirical equilibrium also implicitly assumes that behavior when the mechanism is operated will approximate a Nash equilibrium. On the one hand, though not universally observed, there is evidence that  frequencies of play in many games move towards best responses. For instance, it is common that logistic QRE parameter estimates increase toward best responses when estimated in experiments that are run in multiple periods without repeated game effects \citep{mckelvey:95geb}. On the other hand, one can also see our analysis of mechanisms based on empirical equilibrium as a departure from the Nash equilibrium based analysis that makes it more realistic without losing the regularity and power provided by the assumption that individual incentives are eventually in equilibrium.

It is worth noting that not all empirical evidence supports the assumption of weak payoff monotonicity. \citet{Velez-Brown-2019-SP} documented violations of this property in the second price auction experiments of \citet{Andreoni-Che-Kim-2007-GEB} and the pivotal mechanism experiment of \citet{Cason-et-al-2006-GEB}. In our partnership dissolution environment there is a positive association between the empirical expected payoff of an action and the probability with which it is chosen \citep{Brown-Velez-2017-OAU}. Rounding bids to multiples of five, a behavioral regularity observed in these games, can easily induce violations of payoff monotonicity, however.

Thus, we should see weak payoff monotonicity as a regular expression of the positive association between choice and expected utility in a game. We should not see this property as a bullet proof characterization of behavior in games. Consistently, we should verify whether the conclusions we obtained from our analysis survive for continuous variations of it for which there is more empirical support. The following parametric form of the axiom allows us to do this.

\begin{definition}\rm Let $m\in [0,1]$. A profile of strategies $\sigma:=(\sigma_i)_{i\in N}$ is \textit{$m$-weakly payoff monotone for $\Gamma$} if for each $i\in N$ and each pair of actions $\{a_i,\hat a_i\}\subseteq A_i$ such that $U_i(\sigma_{-i},a_i)\geq U_i(\sigma_{-i},\hat a_i)$, we have that $\sigma_i(a_i)\geq m\sigma_i(\hat a_i)$.
\end{definition}

For $m=1$ the property exactly corresponds to weak payoff monotonicity. For $m>n$, $m$-weak payoff monotonicity implies $n$-weak payoff monotonicity. For $m=0$ the property imposes no restrictions in data.

\begin{definition}\rm Let $m>0$. An \textit{$m$-empirical equilibrium of $\Gamma$} is a Nash equilibrium of $\Gamma$ that is the limit of a  sequence of $m$-weakly payoff monotone strategies for $\Gamma$.
\end{definition}

One can prove that for any $m>0$, the  $m$-empirical equilibrium outcome correspondence violates invariance under Maskin monotonic transformations. Moreover, for values of $m$ close enough to one, the $m$-empirical equilibria of WB and LB are separated on the left half and the right half of the Nash range. It is an interesting open question to determine the lowest $m$ for which this is so.

Finally, we conducted empirical equilibrium analysis under the assumption of risk neutrality. This can be seen as the choice of a model in which we can constructively arrive at the main conclusion from our study, i.e., that mechanisms that are essentially equivalent with respect to the Nash equilibrium prediction may be biased towards certain populations. The interpretation of empirical equilibrium analysis as pointing to particular mechanisms to implement certain normative objectives (Sec.~\ref{Sec-implementation}), does assume that the social planner only entertains agents have this type of risk preferences. It would be interesting, but outside of the scope of this paper, to identify mechanisms that implement in empirical equilibria ordinal social choice correspondences independently of risk preferences \citep[c.f.,][]{MEZZETTI-RENOU-2012}.
\section{Conclussion}\label{Sec:summary}
We advanced empirical equilibrium analysis of EPs. Our results reveal that as long as empirical distributions of play in the games induced by these mechanisms are weakly payoff monotone, they can approximate mutual best responses only if they exhibit a form of bias. WB favors the higher valuation agent and LB favors the lower valuation agent. This implies that an arbitrator who uses an EPA within a legal system in which this type of affirmative action is forbidden, can be subject to a legitimate challenge supported by theory and empirical data. Indeed, our results produce a series of comparative statics that are supported by experimental evidence \citep[see our companion paper,][]{Brown-Velez-2017-OAU}.

Our analysis also brings news to the abstract mechanism design paradigm. We learned that an arbitrator can abide by a principle of equity and at the same time exercise a form of affirmative action that guarantees a special treatment for either low or high value agents. This shows that a mechanism designer who accounts for empirical plausibility of equilibria is not constrained by typical invariance properties.

\section{Appendix}\label{Sec:appendix}
For $x\in\R$, $\lfloor x\rfloor$ denotes the floor of $x$, i.e., the greatest integer that is less than or equal to $x$; $\lceil x\rceil$ denotes the ceiling of $x$, i.e., the smallest integer that is greater than or equal to $x$.

\begin{definition}\rm $\sigma:=(\sigma_i)_{i\in N}$ is  \textit{payoff monotone for $(M,\varphi,v)$} if for each $i\in N$ and each pair $\{m_i,n_i\}\subseteq M_i$, $U_\varphi({m_i}|\sigma_{-i};v_i)\geq U_\varphi({n_i}|\sigma_{-i};v_i)$ if and only if $\sigma_i(m_i)\geq \sigma_i(n_i)$.
\end{definition}

The following theorem is useful in our characterization of empirical equilibria of EPs.

\begin{theorem}[\citealp{Velez-Brown-2018-EE}]\rm\label{Th:interior-charact-VB}Let $(M,\varphi)$ be a EPA and $v\in\Theta$. Then, $\sigma$ is an empirical equilibrium of $(M,\varphi,v)$ if and only if it is a Nash equilibrium of $(M,\varphi,v)$ and it is the limit of interior payoff monotone distributions for $(M,\varphi,v)$.
\end{theorem}

The logistic quantal response function with parameter $\lambda\geq0$, denoted by $l^\lambda$, assigns to each $m_i\in M_i$ and each $x\in\R^{M_i}$ the value,
\begin{equation}l^\lambda_{im_i}(x):=\frac{e^{\lambda x_{m_i}}}{\sum_{n_i\in M_i}e^{\lambda x_{n_i}}}.\label{Equation-Logistic-QRE}\end{equation}
It can be easily checked that for each $\lambda>0$, the corresponding logistic quantal response function is continuous and monotone, i.e., for each $x\in \R^{M_i}$ and each pair $\{m,t\}\subseteq M_i$, $x_m\geq x_{t}$ if and only if $l^\lambda_{im}(x)\geq l^\lambda_{it}(x)$.

\begin{proof}[\textbf{Proof of Proposition~\ref{Prop:charc-equili-auctions}}]
In any of the auctions, by bidding $c_h:= v_h/2$ the high-value agent guarantees a payoff at least $v_{h}-c_h=c_h$. By bidding $c_l:= v_l/2$ the low-value agent guarantees a payoff at least $c_l=v_{l}-c_l$. Thus, in a Nash equilibrium the high-value agent's payoff is at least $c_h$ and the low-value agent's payoff is at least $c_l$.

We prove the proposition for the \textit{winner-bid auction}, which we denote by $(M,\varphi)$. The proof for the \textit{loser-bid auction} is symmetric. Let $\gamma$ be the probability with which an agent with the high value gets the object when there is a tie. We prove our proposition for the slightly more general mechanism in which $\gamma\geq1/2$. The alternative tie breaker $\gamma=1$ may be relevant in experimental settings.

Let $\sigma\in N(M,\varphi,v)$ with $v_l<v_h$. Let $p$ be in the support of $\sigma_{h}$. We claim that $p\leq c_h$. Suppose without loss of generality that $p$ is the maximal element in the support of $\sigma_{h}$. Suppose by contradiction that $p\geq c_h+1$. Since $p> c_l$ and $\gamma>0$, the expected payoff of any bid $b>p$ for the low-valuation agent is strictly lower than the expected payoff of $p$. Thus, there is no $b>p$ in the support of $\sigma_l$. Since $p-1\geq c_h$,
\[\begin{array}{rl}U_\varphi({p-1}|\sigma_l;v_h)-U_\varphi({p}|\sigma_l;v_h)\geq &(1-\sigma_l(p)-\sigma_l(p-1))
\\&+\sigma_l(p-1)(\gamma+(1-\gamma)((p-1)-(v_h-p)))
\\&+\sigma_l(p)(p-(v_h-p))
\\=&(1-\sigma_l(p)-\sigma_l(p-1))
\\&+\sigma_l(p-1)(2\gamma-1+2(1-\gamma)(p-c_h))
\\&+2\sigma_l(p)(p-c_h)>0,
\end{array}\]
where the last inequality holds because $\sigma_l$ is a probability distribution, $\gamma\geq1/2$, and $p-c_h\geq 1$. This contradicts $p$ is in the support of $\sigma_h$.

We claim now that $p\geq c_l$. Suppose without loss of generality that $p$ is the minimal element in the support of $\sigma_h$. Suppose by contradiction that $p\leq c_l-1$. We claim that there is no $b<p$ in the support of $\sigma_l$. Suppose by contradiction there is $b<p$ in the support of $\sigma_l$. Since $c_l-p\geq 1$ and $b<p$,
\[U_\varphi({p+1}|\sigma_h;v_l)-U_\varphi({b}|\sigma_h;v_l)\geq\sigma_h(p)(v_l-(p+1)-p)=\sigma_h(p)(2(c_l-p)-1)>0.\]
This contradicts $b$ is in the support of $\sigma_l$. We claim that $\sigma_l(p)=0$. Suppose by contradiction that $\sigma_l(p)>0$. Then,
\[\begin{array}{cl}U_\varphi({p+1}|\sigma_h;v_l)-U_\varphi({p}|\sigma_h;v_l)&\geq \sigma_h(p)\left(\gamma(v_l-(p+1)-p)+(1-\gamma)(-1)\right)\\
&=\sigma_h(p)\left(\gamma(2(c_l-p)-1)-(1-\gamma)\right)\\
&\geq\sigma_h(p)(\gamma-1+\gamma) \geq 0.\end{array}\]

If the inequality above holds strictly, there is a contradiction to $\sigma_l(p)>0$. Since $\sigma_h(p)>0$, $\gamma\geq 1/2$, and $c_l-p\geq 1$, the expression above is equal to zero only when $\gamma=1/2$ and $p=c_l-1$. Suppose then that $\gamma=1/2$ and $p=c_l-1$. Since for each $b<p$, $\sigma_l(b)=0$, we have that
\[\begin{array}{cl}U_\varphi({p+1}|\sigma_l;v_h)-U_\varphi({p}|\sigma_l;v_h)
 &\geq \frac{1}{2}\sigma_l(p)(-1)+\frac{1}{2}\sigma_l(p)(v_h-c_l-(c_l-1))\\
&=\sigma_l(p)(c_h-c_l)>0.
\end{array}\]
This contradicts $\sigma_h(p)>0$. Thus far we have proved that the support of $\sigma_l$ belongs to $(p,+\infty)$. Let $b$ be the minimum element of the support of $\sigma_l$. Thus,  $p<b$. If $b<c_h$, since $\gamma>0$, agent $h$ would benefit by bidding $b$ instead of $p$.  Thus, $b\geq c_h$. Recall that the support of $\sigma_h$ belongs to $(-\infty,c_h]$. Since the expected payoff of $l$ is at least $c_l$, $b\leq c_h$, for otherwise agent $l$, when bidding $b$, would receive the object with probability one and pay the other agent more than $c_h>c_l$. Thus, $b=c_h$. Thus, agent $l$ would benefit by bidding $b-1$ instead of $b$, because $c_h>c_l$ and $\sigma_h(p)>0$. This contradicts $c_h$ is in the support of $\sigma_l$. Thus, the support of $\sigma_l$ is empty. This is a contradiction. Thus, the support of $\sigma_h$ belongs to $[c_l,c_h]$.

Let $p$ be the minimum element of the support of $\sigma_h$ and $b$ an element in the support of $\sigma_l$.  We claim that $b\leq p$. Suppose by contradiction that there is $b>p$ in the support of $\sigma_l$. Since $p\geq c_l$, $b>c_l$. Thus, since $\sigma_h(p)>0$, agent $l$ benefits by bidding $c_l$ instead of $b$. This contradicts $b$ is in the support of $\sigma_l$. Thus, the support of $\sigma_l$ belongs to $(-\infty,p]$.

Finally, let $p$ be the minimum of the support of $\sigma_h$. If $\gamma=1/2$ and $\sigma$ is efficient, the support of $\sigma_l$ belongs to $\{0,...,p-1\}$. Thus, $\sigma_h=\delta_p$, i.e., agent $h$ receives the object and pays $p$ to the other agent. Suppose now that $\sigma$ is inefficient, i.e., $\sigma_l(p)>0$ and $\gamma=1/2$. We claim that $p=c_l$ and $\sigma_l(p)<1/(c_h-c_l)$. Recall that, $p\in[c_l,c_h]$. Since $\sigma_l(p)>0$,
\[\begin{array}{cl}0\geq U_\varphi({p-1}|\sigma_h;v_l)-U_\varphi({p}|\sigma_h;v_l) &=
\sigma_h(p)[p-[(v_l-p)/2+p/2]]
\\&=\sigma_h(p)(p-c_l).
\end{array}\]
Since $\sigma_h(p)>0$, $p=c_l$. Recall that agent $l$ has guaranteed $v_l/2$ in each Nash equilibrium of $(M,\varphi,v)$. If the maximum of the support of $\sigma_l$ is $c_l$, then agent $h$ gets in equilibrium at least what he or she would get by bidding $c_l+1$. This bid gives agent $h$ an expected payoff of $v_h-(v_l/2+1)$. Thus, the aggregate expected payoff is at least $v_h-1$.
\end{proof}

\begin{proof}[\textbf{Proof of Proposition~\ref{Prop:Nash-payoffs}}]We prove the proposition for the winner-bid auction. The proof for the loser-bid auction is symmetric. Let $\sigma\in N(M,\varphi,v)$. By Proposition~\ref{Prop:charc-equili-auctions}, there is $p\in\{v_l/2,...,v_h/2\}$ that is in the support of $\sigma_h$ such that the support of $\sigma_l$ belongs to $\{0,...,p\}$ and the support of $\sigma_h$ belongs to $\{p,...,\overline{p}\}$. Suppose first that $\sigma$ is efficient. Thus, the support of $\sigma_l$ belongs to $\{0,...,p-1\}$. Thus, $\sigma_h=\delta_p$. Thus, $\pi_l(\sigma)=v_l/2+t$ and $\pi_h(\sigma)=v_h/2+ES(v)-t$ for some $t\in\{0,1,...,ES(v)\}$. One can easily see that for each $t\in\{0,1,...,ES(v)\}$ the distributions $\sigma_l=\delta_{p-1}$, $\sigma_h=\delta_p$ with $p=v_l/2+t$ is a Nash equilibrium. 
\end{proof}

\begin{proof}[\textbf{Proof of necessity in Theorem~\ref{Thm:bias1}}]We prove that $h$'s expected payoff in an empirical equilibrium of the winner-bid auction is bounded below by the expressions in the theorem. The proof for the loser-bid auction is symmetric.

Let $(M,\varphi)$ be the winner-bid auction and $v\in\Theta$. Let $\sigma\in N(M,\varphi,v)$. For $i\in N$, let $c_i:= v_i/2$. By Proposition~\ref{Prop:charc-equili-auctions} there is $p\in\{c_l,...,c_h\}$ separating the supports of $\sigma_l$ and $\sigma_h$, i.e., such that the support of  $\sigma_l$ belongs to $\{0,...,p\}$; the support of $\sigma_h$ belongs to $\{p,...,\overline{p}\}$; and $p$ belongs to the support of $\sigma_h$. Suppose that $p\geq c_l+1$. Since $p$ is a best response to $\sigma_l$ for $h$ with type $v_h$,  $h$'s expected payoff of $p$ should be at least the expected payoff of $p-1$. Thus,
\[v_h-p\geq \sigma_l(p-1)\left[\frac{1}{2}(v_h-(p-1))+\frac{1}{2}(p-1)\right]+(1-\sigma_l(p-1))(v_h-(p-1)).\]
Equivalently,
\begin{equation}\label{Eq:Theorem-Eq1}
  \sigma_l(p-1)\geq1/(c_h-p+1).
\end{equation}
Let $\sigma$ be an efficient empirical equilibrium of $(M,\varphi,v)$ and let $p$ be its associated separating bid (in the support of $\sigma_h$). Let $\{\sigma^\lambda\}_{\lambda\in\N}$ be a sequence of interior payoff monotone distributions for $(M,\varphi,v)$ that converges to $\sigma$ as $\lambda\rightarrow\infty$. This sequence exists by Theorem~\ref{Th:interior-charact-VB}.  For  $i\in N$  and  $\{r,d\}\subseteq\{0,1,...,\overline{p}\}$ such that $r\leq d$, let $\Delta_i(r,d)$ be the difference in expected utility for agent $i$ in the \textit{winner-bid auction} between the two situations in which agent $i$ bids strictly to the left of $r$ and bids exactly $d$, conditional on agent $-i$ bidding $r$.\footnote{Note that conditional on agent~$-i$ bidding $r$, the payoff for agent $i$ when bidding strictly to the left of $r$ is exactly $r$. Thus, $\Delta_i(r,d)$ is well defined.} Using this notation we have that when $b<d$,
\begin{equation}U_\varphi(b|\sigma_{-i},v_i)-U_\varphi(d|\sigma_{-i},v_i)=\sum_{r<b}\sigma_{-i}(r)(d-b)+\sigma_{-i}(b)(d-c_i)+\sum_{b<r\leq d}\sigma_{-i}(r)\Delta_i(r,d).\label{Eq:Thm-Eq2}\end{equation}

We prove that the expected payoff for $h$ given $\sigma$ satisfies the lower bound in the statement of the theorem.

\textbf{Case 1}: $ES(v)=1$.  Recall that we require valuations to be positive. Thus, $c_l>0$. Let $\lambda\in\N$. By (\ref{Eq:Thm-Eq2}),
\[\begin{array}{rl}U_\varphi({c_l-1}|\sigma_{h}^\lambda,v_l)-U_\varphi({c_l}|\sigma_{h}^\lambda,v_l)&=
\sum_{r<c_l-1}\sigma_{h}^\lambda(r)+\sigma_h^\lambda(c_l-1)(c_l-c_l)+\sigma_h^\lambda(c_l-1)\Delta_l(c_l,c_l)\\
&=\sum_{r<c_l-1}\sigma_{h}^\lambda(r)+0+0\geq0.\end{array}\]
By payoff monotonicity, $\sigma^\lambda_{l}(c_l)\leq \sigma^\lambda_{l}(c_l-1)$. By convergence, $\sigma_{l}(c_l)\leq \sigma_{l}(c_l-1)$. Thus, $\sigma_l(c_l)\leq 1/2$. We claim that $p=c_l$. Suppose by contradiction that $p=c_h=c_l+1$. By Proposition~\ref{Prop:charc-equili-auctions}, the support of $\sigma_l$ belongs to $\{0,...,c_l\}$. Thus,
\[U_\varphi({c_l}|\sigma_{l},v_h)-U_\varphi({c_h}|\sigma_{l},v_h)=(1-\sigma_l(c_l))(c_h+1)+\sigma_l(c_l)c_h-c_h=1-\sigma_l(c_l)>0.\]
This contradicts $\sigma$ is a Nash equilibrium of $(M,\varphi,v)$. Thus, $p=c_l$. Since $\sigma$ is efficient, $\pi_h=v_h/2+1$. 

\textbf{Case 2}: $ES(v)>1$.  Let $t:= p-c_l-1$ and suppose that $t\geq 1$.  Let $d:= p-1$. Then
\[\Delta_l(d,d)=d-((v_l-d)/2+d/2)=t.\]
Let $r<d$ and $n=d-r$. Then,
\begin{equation}\Delta_l(r,d)=r-(v_l-d)=c_l+t-n-(2c_l-c_l-t)=2t-n.\label{Eq:2t-n}\end{equation}
We claim that for each $\max\{0,c_l-3t\}\leq b<d$ there is $\Lambda\in\N$ such that for each $\lambda\geq \Lambda$
\begin{equation}U_\varphi({b}|\sigma^\lambda_{h},v_l)-U_\varphi({d}|\sigma^\lambda_{h},v_l)\geq 0.\label{Eq:sigma-lambda}\end{equation}
Note that if this claim is true for $b$, then by payoff monotonicity and convergence, $\sigma_l(b)\geq\sigma_l(d)$.

By (\ref{Eq:Thm-Eq2}), for each $\max\{0,c_l-3t\}\leq b<d$, we have that
\begin{equation}\begin{array}{rl}U_\varphi({b}|\sigma_{h}^\lambda,v_l)-U_\varphi({d}|\sigma_{h}^\lambda,v_l)&=
\sum_{r<b}\sigma_h^\lambda(r)(d-b)+\sigma_h^\lambda(b)(d-c_l)
+\sum_{b<r<d}\sigma_h^\lambda(r)\Delta_l(r,d)\\&+\sigma_h^\lambda(d)\Delta_l(d,d)\\
&=
\sum_{r<b}\sigma_h^\lambda(r)(d-b)+\sigma_h^\lambda(b)t
+\sum_{b<r<d}\sigma_h^\lambda(r)\Delta_l(r,d)+\sigma_h^\lambda(d)t\\
&\geq \sum_{b<r<d}\sigma_h^\lambda(r)\Delta_l(r,d).
\end{array}\label{Eq:Thm-Eq3}\end{equation}

Let $c_l-t-1\leq b< d$. By (\ref{Eq:2t-n}), for each $d-2t\leq r$, $\Delta_l(r,d)\geq 0$.  Thus, by (\ref{Eq:Thm-Eq3}), for each $\lambda\in\N$, $U_\varphi({b}|\sigma_{h}^\lambda,v_l)-U_\varphi({d}|\sigma_{h}^\lambda,v_l)\geq 0$. We complete the proof of our claim by proving by induction on $\eta\in\{1,...,2t\}$ that the claim holds for $b=c_l-t-\eta$. Let $\eta\in\{2,...,2t\}$ and suppose that for each $c_l-t-\eta<g<d$, $\sigma_l(g)\geq \sigma_l(d)$. Let $b:= c_l-t-\eta$. We claim that there is $\Lambda\in\N$ for which for each $\lambda\geq \Lambda$,  \[\sum_{b<r< d}\sigma_h^\lambda(r)\Delta_l(r,d)\geq0.\]
Note that
\[\begin{array}{rl}\sum_{b<r<d}\sigma_h^\lambda(r)\Delta_l(r,d)&=
\sum_{c_l-t-\eta+1\leq r\leq c_l-t-1}\sigma_h^\lambda(r)\Delta_l(r,d)+\sigma_h^\lambda(c-t)\Delta_l(c_l-t,d)
\\&+\sum_{c_l-t+1\leq r\leq c_l-t+\eta-1}\sigma_h^\lambda(r)\Delta_l(r,d)+\sum_{c_l-t+\eta\leq r<d}\sigma_h^\lambda(r)\Delta_l(r,d).
\end{array}\]
Since for each $r\geq c_l-t$, $\Delta_l(r,d)\geq 0$,
\[\begin{array}{rl}\sum_{b<r< d}\sigma_h^\lambda(r)\Delta_l(r,d)&\geq \sum_{c_l-t-\eta+1\leq r\leq c_l-t-1}\sigma_h^\lambda(r)\Delta_l(r,d)
\\&+\sum_{c_l-t+1\leq r\leq c_l-t+\eta-1}\sigma_h^\lambda(r)\Delta_l(r,d)
\\
&=\sum_{1\leq r\leq \eta-1}\sigma_h^\lambda(c_l-t-r))(-r)+
\\&\sum_{1\leq r\leq \eta-1}\sigma_{h}^\lambda(c_l-t+r)r
\\
&=
\sum_{1\leq r\leq \eta-1}(\sigma_{h}^\lambda(c_l-t+r)-\sigma_h^\lambda(c_l-t-r))r.\end{array}\]

Thus, in order to prove our claim it is enough to show that there is $\Lambda\in\N$ such that for each $\lambda\geq \Lambda$ and each $r\in\{1,..,\eta-1\}$, $\sigma_{h}^\lambda(c_l-t+r)\geq \sigma_h^\lambda(c_l-t-r)$. Let $r\in\{1,..,\eta-1\}$. Then,
\[\Delta_h(c_l-t+r,c_l-t+r)=c_l-t+r-((v_h-(c_l-t+r))/2+(c_l-t+r)/2)=-(c_h-c_l+t-r).\]
Let $1\leq s\leq 2r-1$. Then,
\[\Delta_h(c_l-t+r-s,c_l-t+r)=c_l-t+r-s-(v_h-(c_l-t+r))=-(2(c_h-c_l+t-r)+s).\]
Observe that
\[\begin{array}{l}U_\varphi({c_l-t+r}|\sigma_{l},v_h)-U_\varphi({c_l-t-r}|\sigma_{l},v_h)=-\sum_{0\leq s<c_l-t-r}\sigma_l(s)2r\\
+\sigma_l(c_l-t-r)(v_h-(c_l-t+r)-((v_h-(c_l-t-r))/2+(c_l-t-r)/2))\\
-\sum_{1\leq s\leq 2r-1}\sigma_l(c_l-t+r-s)\Delta_h(c_l-t+r-s,c_l-t+r)\\
-\sigma_l(c_l-t+r)\Delta_h(c_l-t+r,c_l-t+r).\end{array}\]
Thus,
\[\begin{array}{l}U_\varphi({c_l-t+r}|\sigma_{l},v_h)-U_\varphi({c_l-t-r}|\sigma_{l},v_h)\geq
-(1-\sum_{0\leq s\leq 2r}\sigma_l(c_l-t+r-s))2r\\
+\sigma_l(c_l-t-r)(c_h-c_l+t-r)\\
+\sum_{1\leq s\leq 2r-1}\sigma_l(c_l-t+r-s)(2(c_h-c_l+t-r)+s)\\
+\sigma_l(c_l-t+r)(c_h-c_l+t-r).\end{array}\]
Since $c_l-t-r\geq c_l-t-\eta+1=b+1$, for each $0\leq s\leq 2r$, $b+1\leq c_l-t+r-s$. Since  $r\leq \eta-1$ and $\eta\leq 2t$, $c_l-t+r\leq c_l+t-1<d$. By the induction hypothesis, for each $0\leq s\leq 2r$, $\sigma_l(c_l-t+r-s)\geq \sigma_l(d)=\sigma_l(p-1)$. By (\ref{Eq:Theorem-Eq1}), $\sigma_l(c_l-t+r-s)\geq 1/(c_h-p+1)$. Thus,
\[\begin{array}{l}U_\varphi({c_l-t+r}|\sigma_{l},v_h)-U_\varphi({c_l-t-r}|\sigma_{l},v_h)\geq\\
-(1-\frac{2r+1}{c_h-p+1})2r+\frac{2r}{c_h-p+1}2(c_h-c_l+t-r)>\\
\frac{2r}{c_h-p+1}(2(c_h-c_l)-(c_h-p+1)).\end{array}\]
Since $p\geq c_l$ and $c_h>c_l+1$, $2(c_h-c_l)-(c_h-p+1)>0$. Thus, $U_\varphi({c_l-t+r}|\sigma_{l},v_h)-U_\varphi({c_l-t-r}|\sigma_{l},v_h)>0$. Thus, there is $\Lambda\in\N$ such that for each $\lambda\geq \Lambda$, $U_\varphi({c_l-t+r}|\sigma_{l}^\lambda,v_h)-U_\varphi({c_l-t-r}|\sigma_{l}^\lambda,v_h)>0$. By payoff monotonicity, $\sigma_h^\lambda(c_l-t+r)\geq \sigma_h^\lambda(c_l-t-r)$. Thus, there is $\Lambda\in\N$ such that for each $\lambda\geq \Lambda$ and each $r\in\{1,..,\eta-1\}$, $\sigma_{h}^\lambda(c_l-t+r)\geq \sigma_h^\lambda(c_l-t-r)$. Thus, for each $\lambda\geq \Lambda$, $\sum_{b<r< d}\sigma_h^\lambda(r)\Delta_l(r,d)\geq0$.

In summary, we have proved that when $p-1\geq c_l+1$, for each $\max\{c_l-3t,0\}\leq b\leq p-1$,
\begin{equation}\sigma_l(b)\geq \sigma_l(p-1).\label{Eq-Thm4}\end{equation}
There are two sub-cases.

\textbf{Case 2.1}: $v_l\leq 3v_h/8$.  We claim that $p\leq ES(v)/2-v_l/4+1/2=v_h/4+1/2$. Suppose by contradiction that $p> v_h/4+1/2$. Since $p$ and $c_l$ are integers, $p\geq \lfloor v_h/4+1/2\rfloor+1$ and $c_l\leq \lceil 3v_h/16\rceil\leq 3v_h/16$. Then, $p-1-c_l\geq \lfloor v_h/4+1/2\rfloor+1-\lceil 3v_h/16\rceil$. Since $v_h$ is even, $\lfloor v_h/4+1/2\rfloor\geq v_h/4$ and $p\geq v_h/4+1$. Thus, $p-1-c_l\geq v_h/16$. Thus, $p-1-c_l\geq 1$ whenever $v_h\geq 16$. Since $ES(v)>1$, then $v_h\geq 6$ (recall that we assumed positive even valuations). Direct calculation determines that  $\lfloor v_h/4+1/2\rfloor+1-\lceil 3v_h/16\rceil\geq 1$ for $v_h=\{6,...,14\}$. Since, $t=p-c_l-1\geq v_h/4-c_l\geq v_h/16=c_h/8$,  $c_l-3t\leq c_l-3c_h/8\leq 0$. Thus, $\max\{c_l-3t,0\}=0$. Since $t\geq 1$, by (\ref{Eq-Thm4}), $\sigma_l(p-1)\leq 1/(c_l+t+1)=1/(c_l+p-1-c_l+1)=1/p$. By (\ref{Eq:Theorem-Eq1}), $1/(c_h-p+1)\leq 1/p$. Thus, $p\leq c_h/2+1/2=v_h/4+1/2$. This contradicts $p\geq v_h/4+1$. Since $\sigma$ is efficient and $\sigma_h(p)>0$, the support of $\sigma_l$ belongs to $\{0,...,p-1\}$. Thus, $\pi_h(\sigma)\geq v_h-p\geq v_h-(v_h/4+1/2)=v_h/2+v_h/4-1/2=v_h/2+ES(v)/2+v_l/4-1/2$. Note that if $v_l\leq 3v_h/8$ and $ES(v)=2$, then $v_l=2$. Thus, statements (a) and (b) in the theorem have no overlap.

\textbf{Case 2.2}: $v_l>3v_h/8$.   We claim that $p\leq  v_l/2+(v_h/2-v_l/2)/5+4/5$. Suppose by contradiction that $p>q:= \lfloor v_l/2+(v_h/2-v_l/2)/5+4/5\rfloor$.  Since $v_l/2$ and $v_h/2-v_l/2>1$ are integers, $q\geq v_l/2+1=c_l+1$.  Moreover, $q=v_l/2+(v_h/2-v_l/2)/5+\varepsilon$ where $\varepsilon\in \{0,1/5,2/5,3/5,4/5\}$. Let $n:= p-q\geq1$ and $t_q:= q-c_l\geq 1$. Then, $t_q=v_l/2+(v_h/2-v_l/2)/5+\varepsilon-c_l=(c_h-c_l)/5+\varepsilon$. Thus, $c_l-3t_q= c_l-3(c_h-c_l)/5-3\varepsilon\geq 8(c_l-3c_h/8)/5-3\varepsilon$. Since $v_l>3v_h/8$,  $c_l>3c_h/8$. Since $c_l$ and $c_h$ are integers, $c_l-3c_h/8\geq 1/8$.  Thus, $c_l-3t_q\geq1/5-3\varepsilon$. Thus, if $\varepsilon\leq 1/5$, $c_l-3t_q\geq 0$; if $2/5\leq \varepsilon\leq 3/5$, $c_l-3t_q\geq -1$; and if $\varepsilon=4/5$, $c_l-3t_q\geq -2$. Suppose that $\varepsilon\leq 1/5$. Since $t_q\geq 1$, by (\ref{Eq-Thm4}), $\sigma_l(p-1)\leq 1/(4t_q+1)$. By (\ref{Eq:Theorem-Eq1}), $1/(c_h-p+1)\leq 1/(4t_q+1)$. Thus, $1/(c_h-q-n+1)\leq 1/(4t_q+1)$.  Equivalently, $c_h-4t_q-n\geq q$. Since $n\geq 1$ and $t_q=(c_h-c_l)/5+\varepsilon$, $c_h-4(c_h-c_l)/5-4\varepsilon-1\geq q$. Equivalently, $q-5\varepsilon-1=c_l+(c_h-c_l)/5-5\varepsilon-1\geq q$. This is a contradiction. Suppose that $2/5\varepsilon\leq 3/5$. Since $t_q\geq 1$, by (\ref{Eq-Thm4}), $\sigma_l(p-1)\leq 1/(4t_q-1+1)$. By (\ref{Eq:Theorem-Eq1}), $1/(c_h-p+1)\leq 1/(4t_q)$. Thus, $1/(c_h-q-n+1)\leq 1/(4t_q)$.  Equivalently, $c_h-4t_q-n+1\geq q$. Since $n\geq 1$ and $t_q=(c_h-c_l)/5+\varepsilon$, $c_h-4(c_h-c_l)/5-4\varepsilon\geq q$. Equivalently, $q-5\varepsilon=c_l+(c_h-c_l)/5-5\varepsilon-1\geq q$. This is a contradiction. Finally, suppose that $\varepsilon=4/5$. Since $t_q\geq 1$, by (\ref{Eq-Thm4}), $\sigma_l(p-1)\leq 1/(4t_q-2+1)$. By (\ref{Eq:Theorem-Eq1}), $1/(c_h-p+1)\leq 1/(4t_q-1)$. Thus, $1/(c_h-q-n+1)\leq 1/(4t_q-1)$.  Equivalently, $c_h-4t_q-n+1+1\geq q$. Since $n\geq 1$ and $t_q=(c_h-c_l)/5+\varepsilon$, $c_h-4(c_h-c_l)/5-4\varepsilon+1\geq q$. Equivalently, $q-3=c_l+(c_h-c_l)/5-5\varepsilon+1\geq q$. This is a contradiction. Since $\sigma$ is efficient and $\sigma_h(p)>0$, the support of $\sigma_l$ belongs to $\{0,...,p-1\}$. Thus, $\pi_h(\sigma)\geq v_h-p\geq v_h- ( v_l/2+(v_h/2-v_l/2)/5+4/5)=v_h/2+4ES(v)/5-4/5$.

Suppose now that $ES(v)=2$. Recall that $\sigma$ is efficient. Since $v_l>3v_h/8$, $v_l>2$. Since $\pi_h(\sigma)\geq v_h/2+4ES(v)/5-4/5=v_h/2+4/5$, $p\leq c_l+1$. We prove that $p=c_l$. Suppose by contradiction that $p=c_l+1$, i.e., $\pi_h(\sigma)=v_h/2+1$. By Proposition~\ref{Prop:charc-equili-auctions}, $\sigma_h=\delta_p$. Let $\{\sigma^\lambda\}_{\lambda\in\N}$ be a sequence of interior payoff monotone distributions such that as $\lambda\rightarrow\infty$, $\sigma^\lambda\rightarrow\sigma$. Since $U_\varphi({c_l-2}|\sigma_h;v_l)-U_\varphi({c_l+1}|\sigma_h;v_l)=c_l+1-c_l=1$, there is $\Lambda\in N$ such that for each $\lambda\geq\Lambda$, $U_\varphi({c_l-2}|\sigma^\lambda_h;v_l)>U_\varphi({c_l+1}|\sigma^\lambda_h;v_l)$. By payoff monotonicity, for each $\lambda\geq\Lambda$, $\sigma^\lambda_l(c_l-2)\geq \sigma^\lambda_l(c_l+1)$. By convergence, $\sigma_l(c_l-2)\geq \sigma_l(c_l+1)$. Notice also that for each $\lambda\in\N$, $U_\varphi({c_l-1}|\sigma^\lambda_h;v_l)\geq U_\varphi({c_l}|\sigma^\lambda_h;v_l)$. Thus, for each $\lambda\in\N$, $\sigma_l^\lambda(c_l-1)\geq \sigma_l^\lambda(c_l)$. Now,
\[\begin{array}{rl}U_\varphi({c_l+1}|\sigma^\lambda_l;v_h)-U_\varphi({c_l}|\sigma^\lambda_l;v_h)&=\sum_{b<c_l}\sigma_l^\lambda(b)(-1)+\sigma^\lambda_l(c_l)(c_h+1-c_h)
\\&+\sigma^\lambda_l(c_l+1)(c_h-(c_h-1))
\\&\leq -\sigma^\lambda_l(c_l-2)-\sigma^\lambda_l(c_l-1)\\&+\sigma^\lambda_l(c_l)+\sigma^\lambda_l(c_l+1)\leq 0.\end{array}\]
By payoff monotonicity, for each $\lambda\geq\Lambda$, $\sigma^\lambda_h(c_l)\geq\sigma^\lambda_h(c_l+1)$. By convergence, $\sigma_h(c_l)\geq\sigma_h(c_l+1)$. Since $p=c_l+1$ is in the support of $\sigma_h$, $c_l$ is also in the support of $\sigma_h$. This contradicts that the support of $\sigma_h$ is contained in $\{p,...,\overline{p}\}$.

Finally, suppose that $ES(v)>2$ and $v_l\geq 7v_h/12-7/6$. We claim that $p<v_l/2+(v_h/2-v_l/2)/5+4/5$. Suppose by contradiction that $p=v_l/2+(v_h/2-v_l/2)/5+4/5$. Let $y:= c_l-3(p-1-c_l)$. Direct calculation yields that since $v_l\geq 7v_h/12-7/6$, $y\geq c_h-p$. By (\ref{Eq:sigma-lambda}), there is $\Lambda\in\N$ such that for each $\lambda\geq\Lambda$, and each $\max\{0,c_l-3t\}\leq b<p-1$, $\sigma^\lambda_l(b)\geq \sigma^\lambda_l(p-1)$. Since $p$ is in the support of $\sigma_h$ and $p>c_l$, for each $b<p$, $U_\varphi(b|\sigma_h;v_l)>U_\varphi(p|\sigma_h;v_l)$. Thus, we can suppose without loss of generality that for each $\lambda\geq \Lambda$ and each $b<p$, $U_\varphi(b|\sigma^\lambda_h;v_l)>U_\varphi(p|\sigma^\lambda_h;v_l)$ and consequently $\sigma^\lambda_l(b)\geq \sigma^\lambda_l(p)$. Thus,
\[\begin{array}{rl}U_\varphi(p|\sigma^\lambda_l;v_h)-U_\varphi({p-1}|\sigma^\lambda_l;v_h)&=\sum_{b<p-1}\sigma_l^\lambda(b)(-1)\\
&+\sigma^\lambda_l(p-1)(2c_h-p-c_h)+\sigma^\lambda_l(p)(c_h-p)\\
&=\sigma^\lambda_l(p)(c_h-p)-\sum_{0\leq b<y}\sigma^\lambda_l(b)\\
&-\sum_{y\leq b\leq p-1}\sigma^\lambda_l(b)+(c_h-p+1)\sigma^\lambda_l(p-1)\\
&\leq \sigma^\lambda_l(p)(c_h-p)-\sum_{0\leq b<y}\sigma^\lambda_l(p)\\
&-\sum_{y\leq b\leq p-1}\sigma^\lambda_l(p-1)+(c_h-p+1)\sigma^\lambda_l(p-1)
\\&\leq \sigma^\lambda_l(p-1)(c_h-p+1-4(p-1-c_l)-1)=0,\end{array}\]
where the last equality follows from direct calculation given that $p=c_l+(c_h-c_l)/5+4/5$. By payoff monotonicity, for each $\lambda\geq \Lambda$, $\sigma^\lambda_h(p-1)\geq \sigma^\lambda_h(p)$.  By convergence, $\sigma_h(p-1)\geq \sigma_h(p)$. Thus, $p-1$ is in the support of $\sigma_h$. This contradicts that the support of $\sigma_h$ belongs to $\{p,...,\overline{p}\}$.
\end{proof}

\begin{proof}[\textbf{Proof of sufficiency in Theorem~\ref{Thm:bias1}}]Let $(M,\varphi)$ be an EPA and $v\in\Theta$ such that $v_l<v_h$. We prove that each Nash equilibrium of $(M,\varphi,v)$ satisfying the bounds in the statement of the theorem is an empirical equilibrium. We prove it for WB. The proof for LB is symmetric. We consider five cases for which the same type of construction applies. These cases are exhaustive. Table~\ref{Table:casesThm1} summarizes how they apply to the different statements in the theorem.

\begin{table}
  \centering
  \begin{tabular}{|c|c|c|c|c|c|}
    \hline
    Statement & Case 1 & Case 2 & Case 3 & Case 4 & Case 5 \\\hline
    1 & + &  &  &  &  \\\hline
    2 & + & + & + &  &  \\\hline
    3 & + & + &  & + &  \\\hline
    4 & + & + &  & + & + \\\hline

  \end{tabular}
  \caption{Cases in the proof of sufficiency in Theorem~\ref{Thm:bias1}: If the case in the proof (column) applies to the statement of the theorem (row), there is a $+$ in the corresponding cell.}\label{Table:casesThm1}
\end{table}

\textbf{Case 1}:  Let $(\pi_l,\pi_h)$ be such that $\pi_l=c_l$ and $\pi_h=c_h+ES(v)$. We construct a sequence of weakly payoff monotone distributions that converges to $\sigma$ where $\sigma_l=\delta_{c_l-1}$ and $\sigma_h=\delta_{c_l}$. One can easily see that $\sigma$ is a Nash equilibrium of $(M,\varphi,v)$. Note that for each $b\leq c_l-1$,
\begin{equation}U_\varphi({c_l}|\sigma_l;v_h)-U_\varphi(b|\sigma_l;v_h)\geq U_\varphi({c_l}|\sigma_l;v_h)-U_\varphi(c_{l}-1|\sigma_l;v_h)=v_h-c_l-c_h=c_h-c_l>0,\label{Eq:strictineq1}\end{equation}
and for each $b>c_l$,
\begin{equation}U_\varphi({c_l}|\sigma_l;v_h)-U_\varphi(b|\sigma_l;v_h)=b-c_l>0.\label{Eq:strictineq2}\end{equation}
Thus, $c_l$ is $h$'s unique best response to $\sigma_l$. Note also that for each $b>c_l$,
\begin{equation}U_\varphi({c_l-1}|\sigma_h;v_l)-U_\varphi(b|\sigma_h;v_l)=c_l-(v_l-b)=b-c_l>0.\label{Eq:strictineq3}\end{equation}
Let $\varepsilon>0$ and $t\in\N$. For each $x_h\in \R^{M_h}$ and $x_l\in \R^{M_h}$, let
\[f^{\varepsilon,t}_{h}(x_h)=(1-\varepsilon)\delta_{c_l}+\varepsilon l^t(x),\]
\[f^{\varepsilon,t}_{l}(x_h)=(1-\varepsilon)\delta_{c_l-1}+\varepsilon l^t(x).\]
Let $\{\gamma^t\}_{t\in\N}$ be such that for each $t\in \N$, $\gamma^t$ is a fixed point of the composition of $(f^{\varepsilon,t}_{h},f^{\varepsilon,t}_{l})$ and the expected payoff operator,  i.e., the continuous mapping that assigns to each profile of mixed strategies $\gamma$, \[F^{\varepsilon,t}(\gamma):=(f^{\varepsilon,t}_{l}
(U_\varphi({b}|\gamma_h;v_l)_{b\in\{0,...,\overline{p}\}}),f^{\varepsilon,t}_{h}(U_\varphi({b}|\gamma_l;v_h)_{b\in\{0,...,\overline{p}\}})).\]

Existence of $\gamma^t$ is guaranteed by Brower's fixed point theorem.

We will now describe the intuition of the reminder of the proof of this case. We will spell the details after. The construction of empirical equilibria in other cases will rest partially on arguments that are related to this first case. Observe that as $\varepsilon$ vanishes and $t$ diverges to infinity, the distributions $\gamma^{\varepsilon,t}$ converge to $\sigma$. Thus, we need to show that for appropriately chosen parameters $\varepsilon$ and $t$, $\gamma^{\varepsilon,t}$ is weakly payoff monotone for $v$. We will show that it is indeed payoff monotone for $v$. Note that by continuity, we will be able to select $\varepsilon$ small enough that (\ref{Eq:strictineq1}-\ref{Eq:strictineq3}) hold. Since $\gamma^{\varepsilon,t}$ is interior, $c_l-1$ actually has higher expected utility than $c_l$ for $l$ ($l$ gains one dollar when $h$ bids below $c_l-1$; note that by our assumption $v_l\geq 4$, $c_l-1>0$). To to complete the construction, it is enough to show that one can select $\varepsilon$ and $t$ such that $c_l-1$ has higher utility than any lower bid for $l$ given that $h$ plays $\gamma^{\varepsilon,t}_h$. Note that the difference in utility between $c_l-1$ and any bid below it for $h$ is bounded below by a positive number. This is so because when $h$ bids $c_l-1$ the auction ends in a tie with probability close to one. Bidding below $c_l-1$ guarantees $h$ loses almost for sure. Since the probabilities that $\gamma^{\varepsilon,t}_h$ places in $c_l-1$ and any bid below it are determined by a logistic QRF, the ratio of this probabilities also vanishes, i.e., $h$ plays any bid below $c_l-1$ with much less probability than $c_l-1$. Now, conditional on $h$ bidding $c_l-1$, the difference in utility for $l$ between bidding $c_l-1$ or below this bid is at least one unit. If what $h$ bids below $c_l-1$ is small compared to what she bids $c_l-1$, $l$ prefers to bid $c_l-1$ than any lower bid.

To simplify notation, let  $u_l^{\varepsilon,t}(b):=U_\varphi({b}|\gamma^{\varepsilon,t}_h;v_l)$ and $u_h^{\varepsilon,t}(b):=U_\varphi({b}|\gamma^{\varepsilon,t}_l;v_h)$. For each $t\in\N$,
\[||\sigma-\gamma^{\varepsilon,t}||_{\infty}:=\max_{i\in\{l,h\},b\in \{0,...,\overline{p}\}}|\sigma_i(b)-\gamma^{\varepsilon,t}_i(b)|<\varepsilon.\]
Thus, there is $\eta>0$ such that for each $0<\varepsilon<\eta$, each $t\in\N$, each $b\in\{0,...,\overline{p}\}\setminus\{c_l\}$,
\begin{equation}u_h^{\varepsilon,t}(c_l)-u_h^{\varepsilon,t}(b)>0,\label{Eq:bestresp1}\end{equation}
and for each  $b>c_l$,
\begin{equation}u_l^{\varepsilon,t}(c_l-1)-u_l^{\varepsilon,t}(c_l-1)>0.\label{Eq:bestresp2}\end{equation}
For each $t$ and $\varepsilon$,  $\gamma^{\varepsilon,t}$ is interior. Thus,
\begin{equation}u_l^{\varepsilon,t}(c_l-1)-u_l^{\varepsilon,t}(c_l)=\sum_{b<c_l-1}\gamma^{\varepsilon,t}_h(b)>0.\label{Eq:bestresp3}\end{equation}
By (\ref{Eq:bestresp1}), for each $0<\varepsilon<\eta$ and each $t$, $\gamma^{\varepsilon,t}_h$ is ordinally equivalent to $(u^{\varepsilon,t}_h(b))_{b\in\{0,...,\overline{p}\}}$ (note that $\gamma^{\varepsilon,t}_h$ is determined outside the support of $\sigma_h$ by a logistic QRF).
Let $0<\varepsilon<\min\{\eta,2/(\overline{p}-2)\}$. We claim that we can fin $T\in\N$ such that for each $t\geq T$, $\gamma^{\varepsilon,t}_l$ is ordinally equivalent to $(u^{\varepsilon,t}_l(b))_{b\in\{0,...,\overline{p}\}}$.  By~(\ref{Eq:bestresp2}) and~(\ref{Eq:bestresp3}) it is enough to show that there is $T$ such that for each $t\geq T$, for each $b<c_l-1$, $u_l^{\varepsilon,t}(c_l-1)-u_l^{\varepsilon,t}(b)>0$ (note that $\gamma^{\varepsilon,t}_l$ is determined outside the support of $\sigma_l$ by a logistic QRF).

For each $b<c_l-1$,
\[\begin{array}{rl}u_h^{\varepsilon,t}(c_l-1)-u_h^{\varepsilon,t}(b)=&\sum_{d<b}\gamma^t_l(d)(-(c_l-1-b))+\gamma^t_l(b)(v_h-c_l+1-c_h)\\
&+\sum_{b<d<c_l-1}\gamma^t_l(d)(v_h-c_l+1-d)+\gamma^t_l(c_l-1)(c_h-c_l+1)\\
>&-\varepsilon\overline{p}+2(1-\varepsilon)>0.
\end{array}\]
where the last inequality holds because $\varepsilon<2/(\overline{p}-2)$.

Thus, for each $b<c_{l}-1$.
\[\begin{array}{rl}u_l^{\varepsilon,t}(c_l)-u_l^{\varepsilon,t}(b)=&\sum_{d<b}\gamma^t_h(d)(-(c_l-b))+\gamma^t_h(b)(v_l-c_l-c_l)\\
&+\sum_{b<d<c_l}\gamma^t_h(d)(v_l-c_l-d)+\gamma^t_h(c_l)(c_l-c_l)\\
\geq &-\sum_{d<b}\gamma^t_h(d)c_l+\gamma^t_h( c_l-1)\\
=&\gamma^t_h( c_l-1)\left(1-\sum_{d<b}\frac{\gamma^t_h(d)}{\gamma^t_h( c_l-1)}c_l\right)\\=&
\gamma^t_h( c_l-1)\left(1-\sum_{d<c_l-1}\frac{\gamma^t_h(d)}{\gamma^t_h( c_l-1)}c_l\right)\\
=&\gamma^t_h( c_l-1)\left(1-c_l\sum_{d<c_l-1}\frac{e^{t u_h(d)}}{e^{t u_h(c_l-1)}}c_l\right)\\
=&\gamma^t_h( c_l-1)\left(1-c_l\sum_{d<c_l-1}e^{-t(u_h(c_l-1)-u_h(d))}\right)
\\\geq &\gamma^t_h( c_l-1)\left(1-c_l\sum_{d<c_l-1}e^{-t(-\varepsilon\overline{p}+2(1-\varepsilon))}\right).\end{array}\]
Thus, there is $T\in\N$ such that for each $t\geq T$, $u_l^{\varepsilon,t}(c_l)-u_l^{\varepsilon,t}(b)>0$.

\textbf{Case 2}: $\pi_l=c_l+1$ and $\pi_h=c_h+ES(v)-1$ and $ES(v)>2$. Let $\varepsilon>0$ and $t\in\N$. For each $x_h\in \R^{M_h}$ and $x_l\in \R^{M_h}$, let
\[f^{\varepsilon,t}_{h}(x_h)=(1-\varepsilon)\delta_{c_l+1}+\varepsilon l^t(x).\]
\[f^{\varepsilon,t}_{l}(x_h)=(1/2-\varepsilon)\delta_{c_l-1}+(1/2-\varepsilon)\delta_{c_l}+2\varepsilon l^t(x).\]
Let $\{\gamma^t\}_{t\in\N}$ be such that for each $t\in \N$, $\gamma^t$ is a fixed point of the composition of $(f^{\varepsilon,t}_{h},f^{\varepsilon,t}_{l})$ and the expected payoff operator. As $t\rightarrow\infty$, $\gamma^t_h\rightarrow\delta_{c_l}$ and $\gamma^t_l\rightarrow(1/2)\delta_{c_l-1}+(1/2)\delta_{c_l}$. Let $\sigma$ be this limit distribution profile. Then, $\pi_l(\sigma)=c_l+1$ and $\pi_h(\sigma)=c_h+ES(v)-1$. In order to complete the construction as in Case 1, we need to show that $\varepsilon$ can be selected small enough such that for large $t$, the distribution is strictly monotone with respect to expected payoffs. The expected payoff of $c_l+1$ given $\sigma$ for $h$, is greater than the payoff of any other bid. The expected payoff of $c_l-1$ and $c_l$ given $\sigma$ for $l$ is greater than the payoff of any bid higher than $c_l$. Thus, it is only necessary to show that the expected payoff of $c_l+1$, for the distribution profile $\gamma^t$ for $h$, is eventually greater than the payoff of $c_l$. Note that the expected payoff of $c_l-1$ is never less than that of $c_l$ for $l$. For each $b<c_l-1$,
\[\begin{array}{rl}u_h^{\varepsilon,t}(c_l-1)-u_h^{\varepsilon,t}(b)=&\sum_{d<b}\gamma^t_l(d)(-(c_l-1-b))+\gamma^t_l(b)(v_h-c_l+1-c_h)\\
&+\sum_{b<d<c_l-1}\gamma^t_l(d)(v_h-c_l+1-d)+\gamma^t_l(c_l-1)(c_h-c_l+1)\\
>&-2\varepsilon\overline{p}+4(1/2-\varepsilon).
\end{array}\]
Thus, the argument in Case 1 can be easily reproduced for the sequence of fixed points.

\textbf{Case 3}: $c_l\leq 3c_h/8$, $c_l+1<p\leq v_h/4+1/2$, $\pi_l=p$ and $\pi_h=v_h-p$ and $ES(v)>2$. Let $\varepsilon>0$ and $t\in\N$. For each $x_h\in \R^{M_h}$ and $x_l\in \R^{M_h}$, let
\[f^{\varepsilon,t}_{h}(x_h)=(1-\varepsilon)\delta_{p}+\varepsilon l^t(x).\]
\[f^{\varepsilon,t}_{l}(x_h)=(1/p-\varepsilon)\left(\sum_{b<p}\delta_{b}\right)+(p-1)\varepsilon\delta_{p}+\varepsilon l^t(x).\]
Let $\{\gamma^t\}_{t\in\N}$ be such that for each $t\in \N$, $\gamma^t$ is a fixed point of the composition of $(f^{\varepsilon,t}_{h},f^{\varepsilon,t}_{l})$ and the expected payoff operator. As $t\rightarrow\infty$, $\gamma^t_h\rightarrow\delta_{p}$ and $\gamma^t_l\rightarrow(1/p)\sum_{b<p}\delta_{b}$. Let $\sigma$ be this limit distribution profile. Then, $\pi_l(\sigma)=p$ and $\pi_h(\sigma)=v_h-p$. In order to complete the construction as in Case 1, we need to show that $\varepsilon$ can be selected small enough such that for large  $t$, the distribution is strictly monotone with respect to expected payoffs. The expected payoff of $p$ given $\sigma$ for $h$, is greater than the payoff of any bid greater than $p$. The expected payoff of each bid less than or equal to $p-1$ given $\sigma$ for $l$ is greater than the payoff of any bid greater than or equal to $p$. The expected payoff of $p$ given $\sigma$ for $l$ is greater than the payoff of any bid greater than $p$.  Thus, it is only necessary to show that the expected payoff of $p$, for the distribution profile $\gamma^t$ for $h$, is eventually greater than the payoff of each $b<p$. Since $2\leq c_l+1<p\leq v_h/4+1/2=c_h+1/2$, $(c_h-p+1)/p\geq 1$,
\[\begin{array}{rl}u_h^{\varepsilon,t}(p)-u_h^{\varepsilon,t}(p-1)=&-\sum_{d<p-1}\gamma^t_l(d)+\gamma^t_l(p-1)(v_h-p-c_h)\\
&+\gamma^t_l(p)(c_h-p)\\
>&-(1/p-\varepsilon)(p-1)-\varepsilon+(1/p-\varepsilon)(c_h-p)+(p-1)\varepsilon(c_h-p)
\\=&-1+1/p+\varepsilon(p-1)-\varepsilon+(c_h-p+1)/p-1/p
\\&-\varepsilon(c_h-p)+(p-1)\varepsilon(c_h-p)\geq 0.
\end{array}\]
Moreover, for $b<p$,
\[\begin{array}{rl}u_h^{\varepsilon,t}(b)-u_h^{\varepsilon,t}(b-1)=&-\sum_{d<b-1}\gamma^t_l(d)+\gamma^t_l(b-1)(v_h-b-c_h)\\
&+\gamma^t_l(b)(c_h-b)
\\>&-(1/p-\varepsilon)(b-1)-\varepsilon+2(1/p-\varepsilon)(c_h-b)
\\\geq & -b/p+1/p+\varepsilon(b-1)-\varepsilon+2(c_h-b+1)/p-2/p-2\varepsilon(c_h-b)
\\\geq & 1-2\varepsilon(c_h-b)-\varepsilon-1/p.
\end{array}\]
Thus, the argument in Case 1 can be easily reproduced for the sequence of fixed points.

\textbf{Case 4}:  $v_l>3v_h/8$, $c_l+1<p< v_l/2+(v_h/2-v_l/2)/5+4/5$, $\pi_l=p$ and $\pi_h=v_h-p$ and $ES(v)>2$. Let $y:= c_l-3(p-1-c_l)$. Since $p-1\leq c_l+(v_h/2-v_l/2)/5+4/5-1$ and $c_l>3c_h/8$, $y>0$. Let $n:= p-y=4(p-1-c_l)+1$. Since $p<v_l/2+(v_h/2-v_l/2)/5+4/5$, $1/(4(p-1-c_l)+1)>1/(c_h-p+1)$. Thus, $1/n>1/(c_h-p+1)$. Let $\sigma$ be defined by: $\sigma_h=\delta_p$ and $\sigma_l:= (1/n)\sum_{y\leq b<p}\delta_b$. Then, $U_\varphi(p|\sigma_l;v_h)-U_\varphi({p-1}|\sigma_l;v_h)=-(n-1)/n+(1/n)(c_h-p)>0$. Thus, $p$ is a unique best response to $\sigma_l$ for $h$. Clearly, $\sigma_l$ is a best response to $\sigma_h$ for $l$. Thus, $\sigma$ is a Nash equilibrium with payoffs $\pi_l(\sigma)=p$ and $\pi_h(\sigma)=v_h-p$. Now, for $y\leq b-1<b\leq p-1$,
\[U_\varphi(b|\sigma_l;v_h)-U_\varphi({b-1}|\sigma_l;v_h)\geq -(n-2)/n+(1/n)(c_h-b)+(1/n)(c_h-b)>1.\]
Thus, $U_\varphi(b|\sigma_l;v_h)$ strictly increases in the set $b\in\{y,...,p\}$. Let $r\in\{p,...,\overline{p}\}$ be the maximum for which $U_\varphi(r|\sigma_l;v_h)\geq U_\varphi(y|\sigma_l;v_h)$. Thus, $U_\varphi({r+1}|\sigma_l;v_h)< U_\varphi(y|\sigma_l;v_h)$. Now, for each $b<y$,
\[U_\varphi(y|\sigma_l;v_h)-U_\varphi({b}|\sigma_l;v_h)\geq \sigma_l(y)(c_h-y)\geq ES(v)/n>0.\]
Thus, there is $\varepsilon>0$ for which for each distribution profile $\gamma$ such that $||\sigma-\gamma||_\infty<\varepsilon$, (i) there is a constant $c>0$ such that for each $b\neq p$, $U_\varphi(p|\gamma_l;v_h)-U_\varphi({b}|\gamma_l;v_h)\geq c$; (ii) for each $b\in\{y,...,r\}$ and each $d<y$ or $d>r$, $U_\varphi(b|\gamma_l;v_h)-U_\varphi({d}|\gamma_l;v_h)>0$; (iii) $U_\varphi(b|\gamma_l;v_h)$ strictly increases in the set $b\in\{y,...,p\}$; and (iv) for each $b\leq p-1<d$, $U_\varphi(b|\gamma_h;v_l)-U_\varphi({d}|\gamma_h;v_l)>0$.

Let $t\in\N$ and $\eta:= \varepsilon/2(r-y+2)$.  For each $x_h\in \R^{M_h}$ and $x_l\in \R^{M_h}$, let
\[f^{\varepsilon,t}_{h}(x_h)=(1-\varepsilon/2)\delta_{p}+\eta \sum_{y\leq b\leq r}\delta_b+\eta l^t(x).\]
\[f^{\varepsilon,t}_{l}(x_h)=(1/n-\varepsilon)\left(\sum_{y\leq b<p}\delta_{b}\right)+n\varepsilon l^t(x).\]
Let $\{\gamma^t\}_{t\in\N}$ be such that for each $t\in \N$, $\gamma^t$ is a fixed point of the composition of $(f^{\varepsilon,t}_{h},f^{\varepsilon,t}_{l})$ and the expected payoff operator. Then, $||\gamma^t-\sigma||_{\infty}<\varepsilon$. Thus, for each $t\in \N$, $\gamma^t$ satisfies conditions (i)-(iv) above. By (i) and (ii), for each $t\in N$ and each pair $\{b,d\}\subseteq\{0,...,\overline{p}\}$, $\gamma^t_h(b)\geq \gamma^t_h(d)$ if and only if $U_\varphi(b|\gamma^t_{l};v_h)-U_\varphi({d}|\gamma^t{l};v_h)$. By (iv), for each $t\in N$ and each $b\leq p-1<d$, $U_\varphi(b|\gamma_h^t;v_l)-U_\varphi({d}|\gamma_h^t;v_l)>0$. By (iii) we can reproduce the argument in our proof of necessity in Theorem~\ref{Thm:bias1} and show that for each $t\in \N$ and each $y\leq b<p-1$, $U_\varphi(b|\gamma_h^t;v_l)-U_\varphi({p-1}|\gamma_h^t;v_l)\geq0$. Now, by (i), as $t\rightarrow\infty$, $\gamma^t_h\rightarrow \gamma_h:= (1-\varepsilon/2+\eta)\delta_p+\eta \sum_{y\leq b\leq r}\delta_b$. Direct calculation shows that for each $b<y$, $U_\varphi({p-1}|\gamma_h;v_l)-U_\varphi({b}|\gamma_h;v_l)=(p-1-c_l)\eta$. Thus, there is $T\in N$ such that for each $t\geq T$, if $d<y\leq b\leq p-1$, $U_\varphi(b|\gamma^t_h;v_l)-U_\varphi({d}|\gamma^t_h;v_l)>0$. This implies that for each $i\in N$ and $\{b,d\}\subseteq\{0,...,\overline{p}\}$, $\gamma^t_l(b)\geq \gamma^t_l(d)$ if and only if $U_\varphi(b|\gamma^t_{h};v_l)\geq U_\varphi({d}|\gamma^t{h};v_l)$. Thus, one can complete the proof as in Case 1.

\textbf{Case 5}:  $v_l>3v_h/8$, $v_l< 7v_h/12-7/6$, $c_l+1<p=v_l/2+(v_h/2-v_l/2)/5+4/5$, $\pi_l=p$, $\pi_h=v_h-p$ and $ES(v)>2$. Let $y:= c_l-3(p-1-c_l)$. Let $n=p-1-y+1=4(p-1-c_l)+1$. Since $p-1-c_l\geq 1$, $n\geq 5$. Direct calculation yields,
\begin{equation}n=c_h-p+1.\label{Eq:Suf-case7-1}\end{equation}
Now, $(c_h-p)-y=6((7v_h/12-7/6)-v_l)/5>0$. Since $c_h$, $p$, and $y$ are integers,
\begin{equation}(c_h-p)-y\geq1.\label{Eq:Suf-case7-2}\end{equation}
Let $\sigma_h:=\delta_p$ and $\sigma_l:= (1/n)\sum_{y\leq b\leq p-1}\delta_b$. Clearly, for each $p\leq b<d$, $U_\varphi(b|\sigma_{l};v_h)\geq U_\varphi(d|\sigma_{l};v_h)+1$. Clearly, for each $d<y$, $U_\varphi(y|\sigma_{l};v_h)>U_\varphi(b|\sigma_{l};v_h)$. Let $y\leq d< p-1$. Then,
\[\begin{array}{rl}U_\varphi({d+1}|\sigma_{l};v_h)-U_\varphi(d|\sigma_{l};v_h)&=-(b-y)(1/n)+(1/n)(v_h-(d+1)-c_h)
\\&+(1/n)(c_h-(d+1))
\\&= -(b-y+2)/n+2(c_h-p+1)/n-2/n
\\&\geq 1-2/n>0.\end{array}\]
Let $r\in\{p,...,\overline{p}\}$ be the maximum for which $U_\varphi(r|\sigma_l;v_h)\geq U_\varphi(y|\sigma_l;v_h)$. Thus, for each $y\leq b\leq r$ and $d>r$, $U_\varphi({b}|\sigma_l;v_h)>U_\varphi(d|\sigma_l;v_h)$. Clearly, for each $b<p<d$, $U_\varphi(b|\sigma_{h};v_l)>U_\varphi(p|\sigma_{h};v_l)>U_\varphi(d|\sigma_{h};v_l)$.
Thus, there is $\zeta>0$ such that if $||\gamma-\sigma||_\infty<\zeta$, then
\begin{enumerate}\item[(a)]$U_\varphi({b}|\gamma_{l};v_h)>U_\varphi(d|\gamma_{l};v_h)$ whenever one of the following four conditions is satisfied
(i) $p\leq b<d$;
  (ii) $d<b=y$;
(iii) $y\leq d<p-1$ and $b=d+1$; or (iv) $y\leq b\leq r$ and $d>r$.
\item[(b)] For each $p<d$, $U_\varphi({p}|\gamma_{l};v_h)\geq U_\varphi(d|\gamma_{l};v_h)+1/2$.
\item[(c)]For each $b<p<d$, $U_\varphi(b|\gamma_{h};v_l)>U_\varphi(p|\gamma_{h};v_l)>U_\varphi(d|\gamma_{h};v_l)$.
\end{enumerate}
Let $t\in\N$, $\varepsilon>0$, $\eta:= \varepsilon/2(r-y+2)$, $\tau:= (y+3/2)\varepsilon/n$ such that $2\max\{\varepsilon,\tau\}<\zeta$.  For each $x_h\in \R^{M_h}$ and $x_l\in \R^{M_h}$, let
\[f^{\varepsilon,t}_{h}(x_h)=(1-\varepsilon/2)\delta_{p}+\eta \sum_{y\leq b\leq r}\delta_b+\eta l^t(x).\]
\[f^{\varepsilon,t}_{l}(x_l)=(1/n-\tau)\left(\sum_{y\leq b\leq p-1}\delta_{b}\right)+\varepsilon\left(\sum_{0\leq b\leq y-1}\delta_{b}\right)+\varepsilon\delta_p+(\varepsilon/2) l^t(x).\]
Let $\{\gamma^t\}_{t\in\N}$ be such that for each $t\in \N$, $\gamma^t$ is a fixed point of the composition of $(f^{\varepsilon,t}_{h},f^{\varepsilon,t}_{l})$ and the expected payoff operator. Then, $||\gamma^t-\sigma||_{\infty}<\zeta$. Thus, for each $t\in \N$, $\gamma^t$ satisfies (a)-(c) above. Now,
\[\begin{array}{rl}U_\varphi({p}|\gamma^t_{l};v_h)-U_\varphi({p-1}|\gamma^t_{l};v_h)&=\sum_{0\leq b\leq p-2}\gamma^t_l(b)(-1)\\&+\gamma^t_l(p-1)(v_h-p-c_h)+\gamma^t_l(p)(c_h-p)
\\&\geq -y\varepsilon-(n-1)(1/n-\tau)\\&-\varepsilon/2+(1/n-\tau)(c_h-p)+\varepsilon(c_h-p)\\
& =\varepsilon(c_h-p-y-1/2)\\&+(1/n-\tau)(c_h-p+1-n).\end{array}\]
By (\ref{Eq:Suf-case7-1}) and (\ref{Eq:Suf-case7-2}), $U_\varphi({p}|\gamma^t_{l};v_h)-U_\varphi({p-1}|\gamma^t_{l};v_h)\geq \varepsilon/2>0$. Thus, for each $t\in \N$ and each $\{b,d\}\subseteq\{0,...,\overline{p}\}$, $\gamma^t_h(b)\geq \gamma^t_h(d)$ if and only if $U_\varphi({b}|\gamma^t_{l};v_h)-U_\varphi({d}|\gamma^t_{l};v_h)$. Moreover, since for each $t\in N$ and each $b\neq p$, $U_\varphi({p}|\gamma^t_{l};v_h)\geq U_\varphi({p-1}|\gamma^t_{l};v_h)+\varepsilon/2$, we have that as $t\rightarrow\infty$,
\[\gamma^t_h\rightarrow \gamma_h:= (1-\varepsilon/2+\eta)\delta_{p}+\eta \sum_{y\leq b\leq r}\delta_b.\]
The construction can be completed as in Case 6.
\end{proof}

\begin{proof}[\textbf{Proof of Lemma~\ref{Lem:limits-marriage}}.] $(M,\varphi)$ be a mechanism satisfying the property in the statement of the lemma. Let $v\in\Theta$, such that $v_l<v_h$, $0\leq t\leq t+1\leq ES(v)$. Let $v_l^*:= 2(v_l/2+t)$ and $v_h^*:= v_l^*+2$. Thus, $ES(v^*)=1$ and the set of efficient and equitable allocations for~$v^*$ is that in which agent~$h$ receives the object and the payoff of the lower valuation agent is between $v_l/2+t$ and $v_l/2+t+1$. By the hypothesis of the lemma, there is  $\sigma\in N(M,\varphi,v^*)$ that obtains an efficient allocation for $v^*$, i.e., agent $h$ receives the object for sure, and $v_l/2+t\leq \pi_l(\sigma)\leq v_l/2+t+1$. We claim that $\sigma\in N(M,\varphi,v)$. We prove first that $\sigma_l$ is a best response to $\sigma_h$ for $l$ with type $v_l$. Let $m_l\in M_l$ be in the support of $\sigma_l$ and $m'_l\in M_l$. Then
\[\begin{array}{r}U_\varphi({m_l}|\sigma_h;v_l)=U_\varphi({m_l}|\sigma_h;v_l^*)\geq U_\varphi({m_l'}|\sigma_h;v_l^*)\geq U_\varphi({m_l'}|\sigma_h;v_l).
\end{array}\]
Where the first equality holds because in equilibrium $\sigma$, $l$ never receives the object, so both expressions are the same integrals of expected compensation; the first inequality is the equilibrium condition for $l$ with type $v^*_l$; and the third inequality holds because the expected utility index of an allotment for $l$ with value $v_l$ is less than or equal than the expected utility index of the allotment for $l$ with value $v_l^*$, because $v_l^*\geq v_l$.

Let $m_h\in M_h$ be in the support of $\sigma_h$ and $m'_h\in M_h$. Since $\sigma$ assigns the object to agent~$h$ with certainty, $U_\varphi({m_h}|\sigma_l;v_h)-U_\varphi({m_h}|\sigma_l;v_h^*)=v_h-v^*_h$. Since the utility index of $h$ is invariant when receiving an amount of money and no object, $U_\varphi({m_h'}|\sigma_l;v_h)-U_\varphi({m_h'}|\sigma_l;v_h^*)\leq v_h-v^*_h$. Thus,
\[U_\varphi({m_h'}|\sigma_l;v_h)-U_\varphi({m_h}|\sigma_l;v_h)\leq U_\varphi({m_h'}|\sigma_l;v_h^*)-U_\varphi({m_h}|\sigma_l;v_h^*)\leq 0,\]
where the last inequality is the equilibrium condition for $\sigma$ with type $v^*_h$. Thus, $\sigma_h$ is a best response to $\sigma_l$ for $h$ with type $v_h$.

Thus, $\sigma\in N(M,\varphi,v)$ is efficient and such that the lower valuation agent's payoff is between $v_l/2+t$ and $v_l/2+t+1$.
\end{proof}

\bibliography{ref-EI}

\end{document}